\documentclass[a4paper,USenglish,cleveref,autoref,thm-restate]
{article}
\usepackage[utf8]{inputenc}
\usepackage{fullpage}
\usepackage{microtype}
\usepackage{mathtools,amsmath,cases} \usepackage{amssymb,amsthm}
\usepackage[T1]{fontenc}
\usepackage{graphicx} \usepackage{xspace} 
\usepackage[numbers]{natbib} \usepackage{hyperref,color}
\usepackage[mathlines]{lineno} 
\usepackage{dsfont}
\usepackage{microtype}
\usepackage{mathrsfs}
\usepackage{algorithm}
\usepackage{algpseudocode}
\usepackage{thmtools} 
\usepackage{soul,esvect}
\usepackage{epstopdf}
\usepackage{todonotes}

\graphicspath{{figures/}}

\newtheorem{theorem}{Theorem} 
\newtheorem{lemma}[theorem]{Lemma} 

\newtheorem{claim}[theorem]{Claim}
\newtheorem{corollary}[theorem]{Corollary}
\newtheorem{definition}[theorem]{Definition}
\newtheorem{observation}[theorem]{Observation}

\newcommand{\cp}{\textsc{Cycle Packing }}
\newcommand{\cpp}{\textsc{Cycle Packing}}

\newcommand{\dcp}{$d$\textsc{-Cycle Packing}}
\newcommand{\ocp}{\textsc{Odd Cycle Packing}}
\newcommand{\dbd}[1]{{#1}\textsc{-Bounded Degree Vertex Deletion}}

\newcommand{\lt}{\textsf{LT}}
\newcommand{\interior}{\textsf{int}}
\newcommand{\bd}{\textsf{bd}}

\newcommand{\cl}{\textsf{cl}}

\newcommand{\lv}{\textsf{lv}}

\newcommand{\bdr}[1]{\textsf{bd}({#1})}

\newcommand{\cut}[1]{\textsf{cut}({#1})}

\newcommand{\finerm}{\ensuremath{\mathcal {M}_\textsf{f}}}

\newcommand{\wsqr}[1]{\ensuremath{c_{\textsf{sos}}({#1}})}
\newcommand{\wsum}[1]{\ensuremath{c_\textsf{sum}}(#1)}
\newcommand{\wmax}[1]{\ensuremath{c_\textsf{max}}(#1)}
\newcommand{\ev}[1]{\ensuremath{\textsf{end}(#1)}}

\newcommand{\w}[1]{\widetilde{#1}}
\def\vecsign{\mathchar"017E}
\def\dvecsign{\smash{\stackon[-1.95pt]{\vecsign}{\rotatebox{180}{$\vecsign$}}}}
\def\dvec#1{\def\useanchorwidth{T}\stackon[-4.2pt]{#1}{\,\dvecsign}}

\usepackage{stackengine}
\stackMath

\title{ETH-Tight Algorithm for Cycle Packing on Unit Disk Graphs\footnote{This work was supported by the National Research Foundation of Korea(NRF) grant
funded by the Korea government(MSIT) (No.RS-2023-00209069).}}

\author{Shinwoo An\footnote{Pohang University of Science and Technology, Korea. Email: \href{mailto:shinwooan@postech.ac.kr}{\texttt{shinwooan@postech.ac.kr}}} 
 \hspace{50pt}    Eunjin Oh\footnote{Pohang University of Science and Technology, Korea. Email: \href{mailto:eunjin.oh@postech.ac.kr}{\texttt{{eunjin.oh@postech.ac.kr}}}}}

\date{}
\begin{document}  
\maketitle
\begin{abstract}
		In this paper, we consider the \textsc{Cycle Packing} problem 
		on unit disk graphs defined as follows. Given a unit disk graph $G$ with $n$ vertices
            and an integer $k$,
		the goal is to find a set of $k$ vertex-disjoint cycles of $G$ if it exists.
		Our algorithm runs in time $2^{O(\sqrt{k})}n^{O(1)}$.
		This improves the $2^{O(\sqrt{k}\log k)}n^{O(1)}$-time algorithm
		  by Fomin et al. [SODA 2012, ICALP 2017]. 
		Moreover, our algorithm is optimal assuming the exponential-time hypothesis. 
\end{abstract}

\section{Introduction}
	The \cp problem is a fundamental graph problem defined as follows. Given an undirected graph $G$ and an integer $k$,
	the goal is to check if there is a set of $k$ vertex-disjoint cycles of $G$.
	This problem is NP-hard even for planar graphs. This motivates 
	the study from the viewpoints of parameterized algorithms~\cite{lokshtanov2019packing} and approximation algorithms~\cite{friggstad2011approximability,krivelevich2007approximation}. 
	For approximation algorithms, we wish to approximate the maximum number of vertex-disjoint cycles of $G$ in polynomial time.  
	The best known polynomial time algorithm has approximation factor of $O(\sqrt{\log n})$~\cite{krivelevich2007approximation}.
	This is almost optimal in the sense that it is quasi-NP-hard to approximate the maximum number of
	vertex-disjoint cycles of a graph within a factor of $O(\log^{1/2-\epsilon} n)$
	for any $\epsilon>0$~\cite{krivelevich2007approximation}.
	Several variants also have been considered, for instance, finding a maximum number of vertex-disjoint triangles~\cite{hassin2006approximation}, 
    finding a maximum number of vertex-disjoint odd cycles~\cite{kawarabayashi2010odd},
    finding a maximum number of edge-disjoint cycles~\cite{friggstad2011approximability},
	and finding a maximum number of vertex-disjoint cycles in directed graphs~\cite{friggstad2011approximability}.
	
	\medskip
	In this paper, we study the \cp problem from the viewpoint of parameterized
	algorithms when the parameter $k$ is the number of vertex-disjoint cycles. 
	This problem is one of the first problems studied from the perspective of Parameterized Complexity. 
	By combining the Erd\H{o}s-P\'{o}sa 
    theorem~\cite{downey2013fundamentals} 
	with a $2^{O(tw\log tw)}n^{O(1)}$-time standard dynamic programming algorithm for this problem,
	where $tw$ is the treewidth of the input graph, 
	one can solve the \cp problem in $2^{O({k}\log^2 k)}n^{O(1)}$ time.
	This algorithm was improved recently by Lokshtanov et al.~\cite{lokshtanov2019packing}. 
        They improved the exponent on the running time by a factor of $O(\log\log k)$.  
	As a lower bound, no algorithm for the \cp problem runs in $2^{o(tw\log tw)}n^{O(1)}$ time assuming the 
	exponential-time hypothesis (ETH)~\cite{cygan2011solving}. 
	
	 For several classes of graphs, the \cp problem can be solved significantly faster.
  For planar graphs, Bodlaender et al.~\cite{bodlaender2008linear} 
	presented a $2^{O(\sqrt{k})}n^{O(1)}$-time algorithm, and showed that the \cp problem admits a linear kernel. Also, 
	one can obtain an algorithm with the same time  bound using the framework of Dorn et al.~\cite{dorn2010efficient}. 
	Later, Dorn et al.~\cite{dorn2012catalan} presented a  $2^{O(\sqrt k)}n^{O(1)}$-time algorithm which works on $H$-minor-free graphs for a fixed graph $H$. 
	As a main ingredient, they presented a branch decomposition of a  $H$-minor-free graph which has a certain structure called the \emph{Catalan structure}. This structure allows them to bound the ways a  cycle may cross a cycle separator of an $H$-minor-free graph. 
	Also, there is a subexponential-time parameterized algorithm for the \cp problem for map graphs~\cite{fomin2019decomposition}. 
	These results raise a natural question whether a subexponential-time algorithm for the \cp problem can be obtained for other graph classes. 
	
	\medskip
	In this paper, we focus on \emph{unit disk graphs}.  
	For a set $V$ of points in the plane, the \emph{unit disk graph} $G$ 
	is defined as the undirected graph whose  
	vertices correspond to the points of $V$ such that  
	two vertices are connected by an edge in $G$ if and only if 
	their Euclidean distance is at most one. 
	It can be used as a model for broadcast networks: The points of $V$ represent transmitter-receiver stations with the same
	transmission power.  
	Unit disk graphs have been studied extensively for various algorithmic 
	problems~\cite{cabello2015shortest,chan2019approximate,esperet2023distributed,hong2022parallel,jena2022vertex,kaplan2018routing}. 
	Also, several NP-complete problems have been studied for unit disk graphs (and geometric intersection graphs) from the viewpoint of parameterized algorithms, for example, the \textsc{Steiner Tree}, \textsc{Bipartization}, \textsc{Feedback Vertex Set}, \textsc{Clique}, \textsc{Vertex Cover}, \textsc{Long Path} and \textsc{Cycle Packing} problems~\cite{an2021feedback,bandyapadhyay2022true,bhore2023parameterized,chan2023finding,fomin2021eth,fomin_et_al:LIPIcs:2017:7393,DBLP:conf/soda/FominLS12,lokshtanov2022subexponential, 10.1007/11847250_14}. All problems listed above, except for the \textsc{Steiner Tree} problem, admit subexponential-time parameterized algorithms for unit disk graphs.
 The study of parameterized algorithms for unit disk graphs and geometric intersection graphs is currently a highly active research area in Computational Geometry.
    
        To the best of our knowledge, the study of subexponential-time \emph{parameterized algorithms} for unit disk graphs was initiated by  
        Fomin et al.~\cite{DBLP:conf/soda/FominLS12}. They focused on the \textsc{Feedback Vertex Set}\footnote{Given a graph $G$ and an integer $k$, find a set $F$ of $k$ vertices such that $G-F$ does not have a cycle.} and \textsc{Cycle Packing} problems
        and presented $2^{O({k}^{0.75}\log k)}n^{O(1)}$-time algorithms. 
        Later, they were improved to take 
        $2^{O({\sqrt k}\log k)}n^{O(1)}$ time by~\cite{fomin_et_al:LIPIcs:2017:7393}, and 
        this approach also works for the \textsc{Long Path} and  \textsc{Long Cycle} problems\footnote{Given a graph $G$ and an integer $k$, find a path and a cycle of $G$ with $k$ vertices, respectively.} with the same time bound.
	Since the best known lower bound for all these problems is $2^{\Omega({\sqrt k})}n^{O(1)}$ assuming ETH~\cite{fomin_et_al:LIPIcs:2017:7393}, it is natural to ask if these problems admit ETH-tight algorithms. 
	Recently, this question was answered affirmatively for all problems mentioned above, except for the \cp problem~\cite{an2021feedback,DBLP:conf/compgeom/FominLP0Z20}.
        It seems that the approaches used in~\cite{an2021feedback,DBLP:conf/compgeom/FominLP0Z20} are not sufficient for obtaining a faster algorithm for the \cp problem. After preprocessing, they reduce the original problems for unit disk graphs to the weighted variants of the problems for 
        graphs with treewidth $O(\sqrt k)$, and then they use $2^{O(tw)}n^{O(1)}$-time algorithms for the weighted variants of the problems for a graph with treewidth $tw$. 
	However, no algorithm for the \cp problem runs in $2^{o(tw\log tw)}n^{O(1)}$ time for a  graph with  treewidth $tw$ assuming ETH~\cite{cygan2011solving}. 
	
	%

	\subparagraph{Our Result.}
	In this paper, we present an ETH-tight parameterized algorithm
	for the \cp problem on unit disk graphs with $n$ vertices, which runs in $2^{O(\sqrt{k})}n^{O(1)}$ time. 
  No ETH-tight algorithm even for the non-parameterized version was known prior to this work. 
  In the case of the \textsc{Long Path/Cycle} and  \textsc{Feedback Vertex Set} problems,
  ETH-tight algorithms running in $2^{O(\sqrt n)}$ time were already known~\cite{de2020framework} before
  the ETH-tight parameterized algorithms were presented~\cite{an2021feedback,DBLP:conf/compgeom/FominLP0Z20}. 
    As a tool, we introduce a new recursive decomposition of the plane into regions with $O(1)$ boundary components with respect to 
    a unit disk graph $G$ 
    such that the edges of $G$ crossing the boundary of each region form a small number of cliques. 
    It can be used for other problems such as the non-parameterized version of the \textsc{Odd Cycle Packing} problem,
    and the parameterized versions of the \textsc{$d$-Cycle Packing} and \textsc{$2$-Bounded-Degree Vertex Deletion} problems on unit disk graphs. For details, see Appendix~\ref{sec:application}.
    

    \section{Overview of Our Algorithm}\label{sec:overview} \label{sec:annuluscarvingproof-overview}
    In this section, we give an overview of our algorithm for the \cp problem on unit disk graphs.
    We are given a unit disk graph $G=(V,E)$ along with its geometric representation. Here, each edge of $G$ is drawn as a line segment connecting its endpoints. We do not distinguish a vertex of $G$ and its corresponding point of $\mathbb{R}^2$
    where it lies.     
    Let $\mathcal M$ be a partition of the plane into interior-disjoint squares of diameter one. We call it a \emph{map} of $G$,\footnote{It is a grid in the case of unit disk graphs, but 
    we define a map as general as possible in the appendices.}
    and a square of $\mathcal M$ a \emph{cell} of $\mathcal M$.
    Notice that the subgraph of $G$ induced by $M\cap V$ is a clique. 
    Throughout this paper, we let $\alpha$ be the maximum number of cells of $\mathcal M$ intersected by one edge of $G$.
    Note that $\alpha=O(1)$. 
    For any integer $r$, a cell $M$ of $\mathcal M$ is called an \emph{$r$-neighboring cell} of $M'$ of $\mathcal M$ if a line segment connecting $M$ and $M'$ intersects at most $r$ cells of $\mathcal M$. 
    Note that a cell of $\mathcal M$ has $O(1)$ $\alpha$-neighboring cells. 

    For a region $A\subseteq \mathbb{R}^2$, we use $\partial A$ to denote the boundary of $A$.  
    For a subset $U$ of $V$, we let $G[U]$ be the subgraph of $G$ induced by $U$.
    For convenience, we let $G[V\setminus U]=G\setminus U$.  
    We often use $V(G)$ and $E(G)$ to denote the vertex set of $G$ and the edge set of $G$, respectively. 
    For a set $\Gamma'$ of paths and cycles of $G$,
    we let $\ev{\Gamma'}$ be the set of end points of the \emph{paths} of $\Gamma'$.
    Moreover, we may seen $\Gamma'$ as a set of edges, and we let $V(\Gamma')$ be the set of all vertices of the \emph{paths and cycles} of $\Gamma'$. 
    
    \subsection{Standard Approach and Main Obstacles}    \label{sec:difficulty-overview}
    Let $\Gamma$ be a set of $k$ vertex-disjoint cycles of $G$. 
    Ideally, we want to recursively decompose the plane into smaller regions each consisting of $O(1)$ boundary curves so that ``very few'' edges of $\Gamma$ cross the boundary of each region. Then we want to compute $\Gamma$ using dynamic programming on the recursive decomposition of the plane.
    For each region $A$, we want to guess the set $E'$ of edges of $\Gamma$ crossing the boundary of $A$, and then want to guess the pairing $\mathcal P$ of the vertices of $\ev{E'}\cap A$ such that two vertices of  $\ev{E'}\cap A$ belong to the same pair of $\mathcal P$
    if and only if a path of $\Gamma\setminus E'$ has them as its endpoints. 
    Then for a fixed pair $(E', \mathcal P)$,
    it suffices to compute the maximum number of vertex-disjoint cycles of $\Gamma'$ contained in $A$ over all sets $\Gamma'$ of vertex-disjoint cycles and paths of the subgraph of $G$ induced by the edges fully contained in $A$ and the edges of $E'$ which match the given information. 
    In fact, this is a standard way to deal with this kind of problems. 

    The \cp problem on planar graphs can be solved using this approach~\cite{dorn2010efficient}.
    Given a simple closed curve $\eta$ intersecting a planar graph only at its vertices, 
    consider the maximal paths of the cycles of $\Gamma$ contained in the interior of $\eta$. 
    Clearly, they do not intersect in its drawing, and thus they have a Catalan structure. 
    Thus once the crossing points of the cycles of $\Gamma$ with $\eta$ are fixed, 
    we can enumerate $2^{O(w)}$ pairings one of which is the correct pairing of the endpoints of the paths, where $w$ is 
    the number of crossing points between $\Gamma$ and $\eta$. However, we cannot directly apply this approach to unit disk graphs
    mainly because $G$ has a large clique, and two vertex-disjoint cycles of $G$ cross in their drawings.
    More specifically, we have the following three obstacles. 

    \subparagraph{First issue.} We do not have any tool for recursively decomposing the plane into smaller regions such that the boundary of each region is crossed by a small number of edges of $\Gamma$. 
    De Berg et al.~\cite{de2020framework}
    presented an algorithm for computing a rectangle separating $V$ in a balanced way
    such that the total \emph{clique-weight} of the cells of $\mathcal M$
    crossing the boundary of the rectangle is $O(\sqrt{n})$, where the clique-weight of a cell $M$ is defined as $\log (|M\cap V|+1)$.
    Using this, we can show that the boundary of this rectangle is crossed by $O(\sqrt{n})$ edges of $\Gamma$.
    However, this only works for a single recursion step. If we apply the algorithm by de Berg et al.~\cite{de2020framework}
    recursively, a region we obtained after $d$ steps can have $\Theta (d)$ boundary components.
    Note that $d=\omega(1)$ in the worst case. 
    To handle this, whenever the number of boundary components of a current region exceeds a certain constant, 
    we have to reduce the number of boundary components using a balanced separator separating the boundary components. However, it seems unclear if it is doable 
    using the algorithm by de Berg et al.~\cite{de2020framework} as their separator works for fat objects, but
    the boundary components are not necessarily fat. 
    Another issue is that we need a separator of complexity $O(\sqrt k)$ instead of $O(\sqrt n)$.

    \subparagraph{Second issue.} Even if we have a recursive decomposition of the plane into regions such that the boundary of each region $A$ is crossed by $O(\sqrt k)$ edges of $\Gamma$, we have to guess the number of edges of $\Gamma$ crossing $\partial A$ among all edges of $G$ crossing $\partial A$. Since the number of edges of $G$ crossing $\partial A$ can be $\Theta(n)$, a naive approach gives $n^{O(\sqrt k)}$ candidates for the correct guess. This issue happens also for other problems on unit disk graphs such as
    \textsc{Long Path/Cycle} and \textsc{Feedback Vertex Set}. The previous results on these problems~\cite{an2021feedback,DBLP:conf/compgeom/FominLP0Z20} handle this issue by using the concept of the clique-weight of a cell of $\mathcal M$, which was introduced by de Berg et al.~\cite{de2020framework}. 
    We can handle this issue as they did. 

    \subparagraph{Third issue.} Suppose that we have a recursive decomposition with desired properties, and 
    we have the set $E'$ of $O(\sqrt k)$ edges of $\Gamma$ crossing the boundary of $A$ for each region $A$.
    For convenience, assume that $\partial A$ is connected. 
    The number of all pairings of $\ev{E'}\cap \partial A$ is $2^{O(\sqrt k\log k)}$, which exceeds the desired bound. 
    But not all such pairings can be the correct pairing of a set of maximum-number of cycles of $G$. 
    For a vertex $a$ of $\ev{E'}\cap \partial A$, let $\bar a$ be the first point on $A$ from $a$ along the edge of $E'$ incident to $a$. 
    If no two cycles of $\Gamma$ cross in their drawings, 
    the cyclic order of $\bar a, \bar b, \bar a'$ and $\bar b'$ along $\partial A$ is either
    $\langle \bar a, \bar b, \bar a', \bar b' \rangle$ or 
     $\langle \bar b, \bar a, \bar a', \bar b' \rangle$ 
    for any two pairs $(a,b)$ and $(a',b')$ of $\mathcal P$,  
    where $\mathcal P$ denotes the correct pairing of $\ev{E'}\cap A$.
    That is, $\mathcal P$ is a \emph{non-crossing pairing}. 
    It is known that for a fixed set $P$ (which is indeed $\ev{E'}\cap A$), the number of non-crossing pairings of $P$ 
    is $2^{O(|P|)}$ (which is indeed $2^{O(\sqrt k)}$).
    However, two cycles of $\Gamma$ can cross in general, and thus $\mathcal P$ is not necessarily non-crossing. 

    \subsection{Our Methods}\label{sec:methods-overview}
    We can handle the three issues using the following two main ideas.
    
    \subparagraph{Surface cut decomposition of small clique-weighted width.}
    We handle the first issue by introducing a new decomposition for unit disk graphs, which we call a \emph{surface cut decomposition}. It is a recursive decomposition of the plane into regions (called \emph{pieces}) with $O(1)$ boundary components
    such that each piece has \emph{clique-weight} $O(\sqrt{\ell})$, where $\ell$ denotes the number of vertices of degree at least three in $G$.
    Let $\cut A$ denote the set of edges with at least one endpoint on $A$ that intersect $\partial A$. An edge of $\cut A$ might have both endpoints in $A$.     For an illustration, see Figure~\ref{fig:annulus}. 
     We show that the total clique-weight of the cells of $\mathcal M$ containing the endpoints of 
    the edges of $\cut A$ is small for each piece $A$. 
    The clique-weight of $A$ is defined as the total clique-weight of the 
    cells containing the endpoints of $\cut A$. Recall that the clique-weight of a cell $M$ is defined as $\log (|M\cap V|+1)$.

    To handle the second issue, we show that $O(1)$ vertices contained in $M$ lie on the cycles of $\Gamma'$ for each cell $M$, 
    and every cycle of $\Gamma\setminus \Gamma'$ is a triangle, where $\Gamma'$ is the set of cycles of $\Gamma$ visiting at least 
    two vertices from different cells of $\mathcal M$. We call this property the \emph{bounded packedness property}. 
    Assume that we have a cell $M$ of clique-weight $\omega$.
    It is sufficient to specify the vertices in $M$ appearing in $\Gamma'$. 
    For the other vertices in $M$, we construct a maximum number of 
    triangles. 
    By the bounded packedness property of $\Gamma$, the number of choices of the edges in $M$ appearing in $\Gamma'$ is reduced to $|M\cap V|^{O(1)}=2^{O(|\log (|M\cap V|+1)|}=2^{O(\omega)}$. 
    This also implies that the boundary of each region 
    $A$ is crossed by $O( \omega)$ cut edges. 

    \subparagraph{Deep analysis on the intersection graphs of cycles.}
    We handle the third issue as follows. 
    We choose $\Gamma$ in such a way that it has the minimum number of edges among all possible solutions. 
    Consider a piece $A$ not containing a hole. 
    Suppose that we have the set $E'$ of $O(\sqrt k)$ cut edges of $\Gamma$ crossing the boundary of $A$.
    Due to the bounded packedness property, we can ignore the cycles of $\Gamma$ fully contained in a single cell of $\mathcal M$. Let $\Gamma'$ be the set of the remaining cycles of $\Gamma$. 
    Let $\Pi$ be the set of path components of the subgraph of $\Gamma'$ induced by $V\cap A$. An endpoint of a path $\pi$ of $\Pi$ is incident to a cut edge of $A$ along $\Gamma$.
    We extend the endpoint of $\pi$ along the cut edge until it hits $\partial A$. 
    Let $\bar \Pi$ be the set of the resulting paths. 

\begin{figure}
	\centering
	\includegraphics[width=0.55\textwidth]{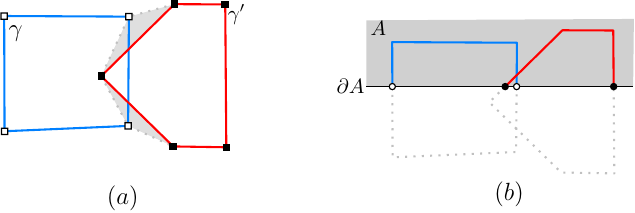}
	\caption{\small 
 (a) The drawings of two cycles of an optimal solution may cross. We cannot replace them into triangles (gray color).
 (b) Two paths of $\bar \Pi$ are cross-ordered, and their drawings cross.       
		\label{fig:crossed-solution}
	}
    \end{figure}

    We aim to compute a small number of pairings of $\ev{\bar \Pi}$ one of which is the correct pairing of $\bar \Pi$. To do this, we consider the intersection graph $\mathcal G$ of the paths of $\bar \Pi$: a vertex of $\mathcal G$ corresponds to a path of $\bar \Pi$, and two vertices are adjacent in $\mathcal G$ if and only if their paths cross in their drawings. We show that the intersection graph $\mathcal G$ of $\bar \Pi$ is $K_{z,z}$-free for a constant $z$. 
    We call this property the \emph{quasi-planar property}. Note that the paths of $\bar \Pi$ may cross even if $\Gamma$ has the bounded packedness property. See Figure~\ref{fig:crossed-solution}.
   
    Then we relate the paring of $\ev{\bar \Pi}$ to $\mathcal G$. 
    We say two paths of $\bar \Pi$, one ending at $a$ and $b$, and one ending at $a'$ and $b'$, are \emph{cross-ordered}, if $a,a',b$ and $b'$ appear along $\partial A$ in this order. 
    For any two \emph{cross-ordered} paths of $\bar \Pi$, their drawings cross since they are contained in $A$.
    Given a pairing $\mathcal P$ of $\ev{\bar \Pi}$, we define the \emph{circular arc crossing graph} of $\mathcal P$ such that each vertex corresponds to a pair of $\mathcal P$, and 
    two vertices are adjacent if their corresponding pairs are cross-ordered. 
    By the previous observation, the circular arc crossing graph is isomorphic to a subgraph of $\mathcal G$. 
    Thus it suffices to enumerate all pairings whose corresponding circular arc crossing graphs are $K_{z,z}$-free. 
    We show that the number of all such pairings is $2^{O(\omega)}$. 

    \medskip 
    In this way, we can enumerate $2^{O(\sqrt{k})}$ pairings of $\ev{\bar\Pi}$ one of which is the correct pairing of $\bar \Pi$
    in the case that $\partial A$ is connected. 
    We can handle the general case where $\partial A$ has more than one curve in a similar manner although there are some technical issues. 
    
    \section{Preliminaries}\label{sec:preliminary}
    For a connected subset $A$ of $\mathbb R^2$, let $\partial A$ denote the boundary of $A$. 
    The \emph{closure} of $A$, denoted by $\cl(A)$, is defined as $A\cup \partial A$.
    Also, the \emph{interior} of $A$, denoted by $\interior(A)$, is defined as $\cl(A)\setminus \partial A$. 
    The \emph{diameter} of $A$ is defined as the maximum Euclidean distance between two points of $A$.
    A \emph{curve} is the image of a continuous function from an unit interval into the plane.
    For any two points $a$ and $b$ in the plane, we call a curve connecting $a$ and $b$ an \emph{$a$-$b$ curve}. 
    A connected subset $C'$ of $C$ is a \emph{subcurve} of $C$.
    For a simple closed curve $C$, $\mathbb{R}^2\setminus C$ consists of two disjoint regions by the boundary curve theorem: an unbounded region and bounded region. 
    We call the unbounded region the $\emph{exterior}$ of $C$, and denote it by $C^\textsf{ex}$. 
    Also, we call the other region the $\emph{interior}$ of $C$, and denote it by $C^\textsf{in}$. 
    If a simple closed curve $C$ intersects a plane graph $G$ only at vertices of $G$,
    we call $C$ a \emph{noose} of $G$. If it is clear from the context, we simply call it a \emph{noose}.

     Let $G=(V,E)$ be a graph.
    For a subset $U$ of $V$, we let $G[U]$ be the subgraph of $G$ induced by $U$.
    For convenience, we say the subgraph of $G$ induced by $V\setminus U$ as $G\setminus U$.  
    We often use $V(G)$ and $E(G)$ to denote the vertex set of $G$ and the edge set of $G$, respectively. By a slight abuse of notation, we let $|G|$ be the number of vertices of $G$. 
    A \emph{drawing} of $G$ is 
    a representation of $G$ in the plane such that the vertices are drawn as points in the plane, and
    the edges are drawn as curves connecting their endpoints. Here, the curves corresponding to two edges can cross. 
    A \emph{drawing} is called a \emph{straight-line} drawing if the edges are drawn as line segments. 
    We sometimes use a graph $G$ and its drawing interchangeably if it is clear from the context.
    We deal with undirected graphs only, and the \emph{length} of a path is defined as the number of edges in the path.
    
    Throughout this paper, for a graph $G=(V,E)$ with vertex-weight $c: V\rightarrow \mathbb R$, we let
\[\wsqr{G}=\sqrt{\sum_{v\in V}(c(v))^2}, \quad  
\wsum{G}=\sum_{v\in V} c(v), \quad  \wmax{G}=\max_{v\in V} c(v).\]

    In this paper, we describe our algorithm in a more general way so that it works for larger classes of geometric intersection graphs such as intersection graphs of similarly-sized disks and squares.
    In particular, our algorithm works on a graph drawn in the plane with a straight-line drawing satisfying the \emph{\textbf{icf}-property}
    and having a \emph{map}.

	%
	

    \subsection{Geometric Tools: ICF-Property and Map Sparsifier} \label{sec:gridspanner}
    Let $G=(V,E)$ be a graph with its straight-line drawing.
    We say two edges \emph{cross} if the drawing of two edges cross. 
    We say $G$ has the \emph{induced-crossing-free} property, the \textbf{icf}-property in short, if for any crossing edges $xx'$ and $yy'$ of $G$, three of $\{x,x',y',y\}$ form a cycle in $G$.
    As an example, a unit disk graph $G$ admits the \textbf{icf}-property.
    \begin{observation} \label{obs:udg-cross}
    For any crossing edges $xx'$ and $yy'$ of a unit disk graph, three of $\{x,x',y',y\}$ form a cycle.
    \end{observation}
    \begin{proof}
        For a point $a,b\in \mathbb R^2$, we denote the Euclidean distance between $a$ and $b$ by $|ab|$.
        Without loss of generality, assume that $|xx'| \geq |yy'|$. 
		Consider the two disks $D$ and $D'$ centered at $x$ and $x'$, respectively, with radius $|xx'|$. 
		Any line segment crossing $xx'$ having their endpoints on the boundary of $D\cap D'$ has length at least $|xx'|$.
		Therefore, either $y$ or $y'$ lies in $D\cap D'$.
		Otherwise, $|yy'| > |xx'|$, which makes a contradiction. 
		Therefore, $x, x'$ and $y$ (or $y'$) form a cycle in a unit disk graph. 
    \end{proof}
 
    From now on, we always assume that $G$ admits the \textbf{icf}-property.
    A subset $M$ of $\mathbb R^2$ is \emph{$r$-fat} if there are two disks $D\subset M\subset D'$ and the radius ratio of $D$ and $D'$ is $r$.
    A family of $r$-fat subsets is called a \emph{similarly-size} family if the ratio of the maximum diameter and minimum diameter among subsets is bounded by a constant $r'$.
    A \emph{map} $\mathcal M$ of $G$ is a partition of a rectangle containing $V(G)$ into a similarly-sized family of convex subsets (called cells) each of complexity $O(1)$ satisfying the conditions \textbf{(M1--M2)}: For each cell $M$ of $\mathcal M$,
    
     \begin{itemize}\setlength\itemsep{-0.1em}
        \item \textbf{(M1)} $V(G)\cap M$ forms a clique in $G$, and 
        \item \textbf{(M2)} each edge of $G$ intersects $O(1)$ cells.
    \end{itemize}
    
    We assume the \emph{general position assumption} that no vertex of $V$ is contained on the boundary of a cell of $\mathcal M$,
    and no two edges of $E$ cross at a point in $\partial M$.
     For two cells $M$ and $M'$, the \emph{distance} between $M$ and $M'$ is the length of the shortest path between them in the dual graph of $\mathcal M$. We write the distance by $d(M,M')$.
    For an integer $\ell> 0$, a cell $M$ is called an \emph{$\ell$-neighbor} of a cell $M'$ if $d(M,M')\leq \ell$.
    Each cell has $O_\ell(1)$ $\ell$-neighbors for any integer $\ell>0$ since the cells are similarly-sized and fat.  
    For a point $v$ in the plane, we use $M_v$ to denote the cell of $\mathcal M$ containing $v$.
   Throughout this paper, we let $\alpha$ be the maximum number of cells of $\mathcal M$ intersected by one edge of $G$.
    Since an edge of $G$ has length at most one, $\alpha$ 
    is a constant.
    
    \begin{definition}[Clique-weight]\label{def:clique-weight}
    For a graph $G=(V,E)$ with the \textbf{icf}-property and a cell $M$ of $\mathcal M$, 
    the \emph{clique-weight} of $M$ is defined as $\log(|V\cap M|+1)$.
    \end{definition}

    

\begin{figure}
	\centering
	\includegraphics[width=0.7\textwidth]{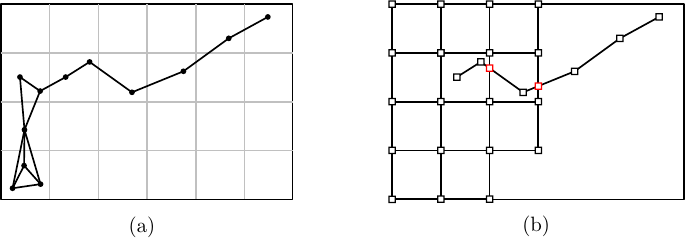}
	\caption{\small (a) Illustration of $G$ and $\mathcal M$.
 (b) The base vertices are marked with black boxes, and the cross vertices are marked with red boxes.        
		\label{fig:map-sparsifier}
	}
    \end{figure}

    Suppose a map $\mathcal M$ is given.
    We define the \emph{map sparsifier} $H$ of $G$ as follows. See Figure~\ref{fig:map-sparsifier}. 
    Consider all $\alpha$-neighboring cells of the cells containing vertices of $G$ of degree at least three. Let $H'$ be the plane graph consisting of all boundary edges of such cells. 
    In addition to them, we add the edges of $G$ whose both endpoints have degree at most two to $H'$.     
    The vertices of $H'$ are called the \emph{base vertices}. 
    It is possible that more than one base vertices are contained in a single cell,
    but the number of base vertices contained in a single cell is $O(1)$. 
    Two edges of $H'$ can cross. In this case, 
    we add such a crossing point as a vertex of $H'$ and split the two edges with respect to the new vertex. 
    These vertices are called the \emph{cross vertices}. Let $H$ be the resulting planar graph, and we call it a \emph{map sparsifier} of $G$
    with respect to $\mathcal M$. 
     We can compute a map sparsifier $H$ and its drawing in polynomial time.   

     \begin{lemma}\label{lem:map-spanner-complexity}
         The number of vertices of $H$ is $O(|V(G)|)$. Among them, at most $O(\ell)$ vertices have degree at least three in $H$,
         where $\ell$ denotes the number of vertices of $G$ of degree $\geq 3$.
     \end{lemma}
     \begin{proof}
       A vertex of $H$ is either a degree-2 vertex of $G$, a corner of a cell of $\mathcal M$, or a cross vertex.      
       The number of vertices of the first type is at most $|V(G)|$,
       and the number of vertices of the second type is $O(\ell)$. This is because
        such a vertex is a corner of an $O(1)$-neighboring cell
         of a vertex of $G$ of degree at least three. 
         Then we analyze the number of cross vertices. 
         No two edges from the boundaries of the cells of $\mathcal P$ cross.
         Also, note that no two edges of $G$ whose both endpoints have degree at most two 
         cross by the \textbf{icf}-property.
         Therefore, a cross vertex $v$ is a crossing point between an edge $e$ of $G$
         and the boundary of a cell $M$ of $\mathcal M$.
         Notice that $M$ is an $O(1)$-neighboring cell of a vertex of degree at least three of $G$,
         and thus the endpoints of $e$ are also contained in $O(1)$-neighboring cells of a vertex of degree at least three. Thus there are $O(\ell)$ edges of $G$ inducing cross vertices of $H$. Since each such edge intersects $O(1)$ cells,     the number of cross edges 
         is $O(\ell)$.                  
         
         To analyze the number of vertices of degree at least three, observe that 
         a vertex of the first type has degree two in $H$, and the number of vertices of the second and third types is $O(\ell)$. Therefore, the lemma holds. 
     \end{proof}

    \subsection{Surface Decomposition and Surface Cut Decomposition.} \label{sec:annulusandac}
    In this subsection, we introduce the concepts of a \emph{surface decomposition} and a \emph{surface cut decomposition},
    which are variants of the branch decomposition and carving decomposition with certain properties, respectively. 
    A point set $A$ is \emph{regular closed} if the closure of the interior of $A$ is $A$ itself. 
    In this paper, we say a regular closed and interior-connected region $A$ is an \emph{$s$-piece} 
    if $\mathbb{R}^2\setminus A$ has $s$ connected components. If $s=O(1)$, we simply say $A$ a \emph{piece}. Note that a piece is a planar surface, and this is why we call the following decompositions the \emph{surface} decomposition and \emph{surface} cut decomposition. 
    In this case, we call $s$ the \emph{rank} of $A$. 
    The boundary of each component of $\mathbb{R}^2\setminus A$ is a closed curve. We call such a boundary component a \emph{boundary curve} of $A$.\footnote{Because of computational issues, a boundary curve of a piece we consider in this paper will be a polygonal curve of complexity $O(n)$.}

        \begin{lemma}
        For a piece $A$, the boundary of each connected component of $\mathbb{R}^2\setminus A$ is a closed curve.
        Moreover, two boundary curves intersect at most once at a single point. 
        \end{lemma}
        \begin{proof}
        For the first statement, $\partial A$ divides the plane into $\interior(A)$ and the \emph{holes} of $A$.
        Here, a hole is a connected component of $\mathbb R^2\setminus A$. Note that a hole is a non-empty connected open set, and thus its boundary is a closed curve.          
        Next, assume $\partial F_1 \cap \partial F_2$ consists of at least two connected components for two holes $F_1$ and $F_2$ of $A$. 
        Then $\partial F_1\cup \partial F_2$ divides the plane into at least four faces, two of them are contained in $\interior(A)$. These two faces must be connected since $A$ is interior-disjoint, which makes a contradiction.
        Now assume that $\partial F_1 \cap \partial F_2$ consists of a single connected component containing at least two points.
        That is, $\partial F_1 \cap \partial F_2$ is a curve. 
        Since $\partial F_1 \cap \partial F_2$ is not contained in $F_1$ (and $F_2$), it must be contained in $A$. 
        On the other hand, it is not contained in the interior of $A$, and thus it is not contained in the closure of $A$ since it is a curve.
        This contradicts that $A$ is regular-closed. Therefore, 
        two boundary curves intersect at most once at a single point.        
        \end{proof}
        
 \subparagraph{Surface decomposition.} \label{sec:annulus-decomposition}
        A \emph{surface decomposition} of a plane graph $H$ with vertex weights $c: V(H)\rightarrow \mathbb{R}^+$ is 
        a pair $(T, \mathcal A)$ 
        where $T$ is a rooted binary tree, and 
         $\mathcal A$ is a mapping that maps a node $t$ of $T$ into a piece $A_t$ in the plane satisfying the conditions \textbf{(A1--A3)}.
        For a node $t$ and two children $t',t''$ of $t$,

        \begin{itemize}\setlength\itemsep{-0.1em}
        \item \textbf{(A1)} $\partial A_t$ intersects the planar drawing of $H$ only at its vertices,    
	  \item \textbf{(A2)} $A_{t'}, A_{t''}$ are interior-disjoint and $A_{t'}\cup A_{t''}=A_t$, and
        \item \textbf{(A3)} $|V(H)\cap A_t|\leq 2$ if $t$ is a leaf node of $T$.
	\end{itemize}

\begin{figure}
	\centering
	\includegraphics[width=0.85\textwidth]{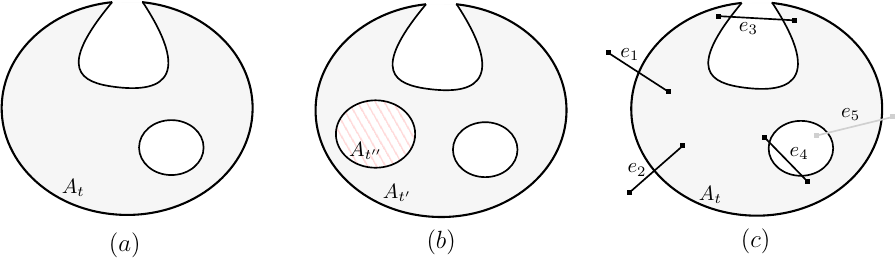}
	\caption{\small (a) A planar surface with rank 2. 
        (b) Two regions $A_{t'}$ and $A_{t''}$ partition $A_t$.
        (c) The sold edges ($e_1,e_2, e_3$ and $e_4$) are cut edges, and $e_5$ is not a cut edge of $A_t$.
		\label{fig:annulus}
	}
    \end{figure}

    For an illustration, see Figure~\ref{fig:annulus}(a--b). 
    The \emph{weight} of a node $t$ is defined as the sum of weights of the vertices of $V(H)$ lying on $\partial A_t$. 
    The \emph{weighted width} of $(T,\mathcal A)$ is defined as the maximum weight of the nodes of $T$.
    We will prove the following theorem in Sections~\ref{sec:cycle-separator} and \ref{sec:annulus}. 
    \begin{restatable}{theorem}{annulus}
    \label{thm:annulus}
        For a  plane graph $H=(V,E)$ with vertex weight $c(\cdot)$ with $1\leq c(v)\leq n^{O(1)}$ for all $v\in V$, 
	one can compute a surface decomposition of weighted width $O(\sqrt{\sum_{v\in V}(c(v))^2})$ in $O(n\log n)$ time,
		where $n$ denotes the number of vertices of $H$.
    \end{restatable}

    \subparagraph{Surface cut decomposition.} \label{sec:ac}
    A key idea of our result lies in the introduction of a surface cut decomposition. Let $G$ be an undirected graph drawn in the plane, which is not necessarily planar. 
    We recursively decompose the plane into pieces 
    such that the number of edges of $E$ crossing the boundary of each piece $A$ is small. Once this is done, the number of all possible cases for the interaction between the parts of $G$ contained in $\interior(A)$ and $\mathbb{R}^2\setminus\interior(A)$ is small, and thus for each possible case, we can break the problem into two subproblems, one for $\interior(A)$ and one for $\mathbb{R}^2\setminus\interior(A)$.
        A \emph{surface cut decomposition}, sc-decomposition in short, of $G$ is defined as a pair $(T, \mathcal A)$ 
        where $T$ is a rooted binary tree, and $\mathcal A$ is a mapping that maps a node $t$ of $T$ into a piece $A_t$ in the plane satisfying the conditions \textbf{(C1--C4)}.
    Let $\mathcal M$ be a given map. 
    For a node $t$ of $T$ and two children $t',t''$ and a cell $M$ of $\mathcal M$,
    
        \begin{itemize}\setlength\itemsep{-0.1em}
        \item \textbf{(C1)} $V(G)\cap \partial A_t=\emptyset$, 
    \item \textbf{(C2)}  $A_{t'}, A_{t''}$ are interior-disjoint, $A_{t'}\cup A_{t''}=A_t$, 
        \item \textbf{(C3)} $V(G)\cap A_t$ is contained in the union of at most two cells of $\mathcal M$ for a leaf node $t$ of $T$,
               \item \textbf{(C4)} 
    there are $O(1)$ leaf nodes of $T$ containing points of $M\cap V(G)$ in their pieces. 
    \end{itemize}

    Notice that a main difference between the surface decomposition and the surface cut decomposition is the condition \textbf{(C1)}.
    For dynamic programming algorithms, we will define subproblems for the subgraph $G_t$ of $G$ induced by $V\cap A_t$. Thus it is sufficient to care about the edges incident to the vertices in $G_t$. 
    Let $\cut t$ be the set of edges of $G$ with at least one endpoint in $A_t$ 
    intersecting $\partial A_t$. 
    Notice that an edge in $\cut t$ might have both endpoints in $A_t$.
    See Figure~\ref{fig:annulus}(c). 

    The \emph{clique-weighted width} of a node $t$ (\emph{with respect to} $\mathcal M$) is defined as the sum of the
    clique-weights of 
    the cells of $\mathcal M$ containing 
    the endpoints of the edges of $\cut t$. 
    The \emph{clique-weighted width} of an sc-decomposition $(T, \mathcal A)$ (\emph{with respect to} $\mathcal M$) is defined as the 
    maximum clique-weight of the nodes of $T$.
    If we can utilize two geometric tools from Section~\ref{sec:gridspanner},
    we can compute an sc-decomposition of small width with respect to $\mathcal M$. 
    We will prove the following theorem and corollary in Section~\ref{sec:annuluscarving}.
    



    \begin{restatable}{theorem}{annuluscarving} \label{thm:annuluscarving}
        Let $G$ be a graph admits \textbf{icf}-property. 
        Given a map $\mathcal M$ of $G$,
        one can compute an sc-decomposition of clique-weighted width $O(\sqrt{\ell})$
        in polynomial time, where
        $\ell$ is the number of vertices of degree at least three in $G$.  
    \end{restatable}


    A unit disk graph admits the \textbf{icf}-property, and a grid that partitions the plane into axis-parallel squares of diameter one is a map of a unit disk graph.

    \begin{restatable}{corollary}{carvingudg} \label{thm:carvingudg}
        Let $G=(V,E)$ be a unit disk graph and $\ell$ be the number of vertices of degree at least three in $G$.
        If the geometric representation of $G$ is given, one can compute an sc-decomposition of clique-weighted width $O(\sqrt \ell)$ in polynomial time. 
    \end{restatable}

\section{Weighted Cycle Separator of a Planar Graph}\label{sec:cycle-separator}
Let $H$ be a triangulated plane graph such that each vertex $v$ has \emph{cycle-weight} $c(v)$ 
and 
\emph{balance-weight} $b(v)$ with $1\leq c(v)\leq |H|^{O(1)}$ and $0\leq b(v)$. 
For a constant $0<\alpha<1$,
a subset $S$ of $V(H)$ is called an $\alpha$-\emph{balanced separator} of $H$ 
if the total balance-weight of each connected component of $H \setminus S$ is
at most $\alpha$ of the total balance-weight of $H$. 
Moreover, we call $S$ a $\emph{cycle separator}$ if $S$ forms a simple cycle in $H$.
In this section, we show that 
$H$ has a $8/9$-balanced cycle separator of weight $O\left(\sqrt{\sum_{v\in V(H)}(c(v))^2}\right)$.
This generalizes the results presented in~\cite{djidjev1997weighted} and in~\cite{harpeled17cycleseparator}. More specifically, 
Djidjev~\cite{djidjev1997weighted} showed that $H$ has a $2/3$-balanced separator with the desired cycle-weight, but the separator is not necessarily a cycle. 
On the other hand, Har-Peled and Nayyeri~\cite{harpeled17cycleseparator} showed that 
$H$ has a $2/3$-balanced cycle separator of the desired cycle-weight if the cycle-weight and balance-weight of every vertex are equal to one. 

\subsection{Balanced Cycle Separator}
We first compute a $2/3$-balanced cycle separator $S$, which might have a large cycle-weight.
To do this, we use the \emph{level tree} of $H$ introduced by Lipton and Tarjan~\cite{Lipton80planarseparator}.
We choose the root vertex $r$ of $V(H)$ arbitrary.
The \emph{cycle-weight} of a simple path in $H$ is defined as the 
sum of the cycle-weights of all vertices of the path.
Then the \emph{level} of a vertex $p$ of $H$, denoted by $\lv(p)$, is defined as 
the minimum cycle-weight of an $r$-$p$ path in $H$. 
The \emph{level tree} is defined as the 
minimum cycle-weight path tree, that is, 
each $r$-$p$ path of the tree has minimum cycle-weights among all $r$-$p$ paths of $H$.
We denote the level tree by $\lt(H)$.
A balanced cycle separator can be obtained from $\lt(H)$ but this separator might have a large cycle-weight.
\begin{lemma}[{\cite{lipton1979separator}}] \label{lem:largecyclesep}
	Given a triangulated planar graph $H$, we can find a root vertex $r$ of $H$ and an edge  $(u,v)\in E(H)$ in $O(|V|)$ time such that
	\begin{itemize}\setlength\itemsep{-0.1em}
		\item $\pi_u$ and $\pi_v$ are edge-disjoint, 
		\item the cycle $S=\pi_u\cup \pi_v\cup (u,v)$ is a $2/3$-balanced cycle separator of $H$,
	\end{itemize}
	where $\pi_u$ is an $r$-$u$ path of the level tree $\lt(H)$ of $H$ rooted at $r$.
\end{lemma}

In the rest of the section, $r, u,v$ and $S$ are the root, two vertices and the cycle separator specified by Lemma~\ref{lem:largecyclesep}, respectively.
The maximum level of the vertices in $S$ is either $\lv(u)$ or $\lv(v)$.
Without loss of generality, we assume that $\ell_\textsf{max}=\lv(u)$ is the maximum level.
Also, the minimum level of the vertices in $S$
is the level of $r$. We let $\ell_\textsf{min} = \lv(r)=c(r)$.
We use $\textsf{parent}(w)$ to denote the parent node of $w$ in $\lt(H)$.

\subsection{Cycle Separators with Small Cycle-Weight}
\begin{figure}
	\centering
	\includegraphics[width=0.8\textwidth]{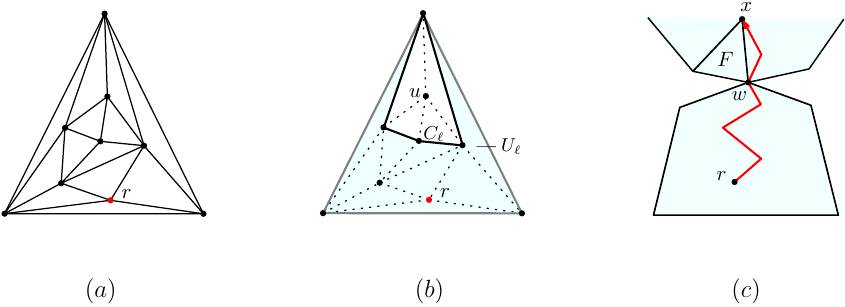}
	\caption{\small 
		(a) Triangulated planar graph $H$ with the root node $r$. 
        (b) Illustrates $U_\ell$ by the blue region, and $C_\ell$ by the thick black curve.
		(c) An $r$-$x$ path contained in $\interior(U_\ell)$ encounters a cut vertex $w$.
		\label{fig:simpcycc}
	}
\end{figure}

In this subsection, we construct a sequence of {vertex-disjoint} cycles each of which crosses $S$ exactly
twice and has small cycle-weight.
Let $c^*=\sqrt{\sum_{v\in V(H)}(c(v))^2}$ be the desired weight. Note that $c(v)\leq c^*$ for any vertex $v\in H$. 
If the cycle-weight of $S$ is at most $8c^*$, $S$ is a desired balanced cycle separator.
Thus we assume that the cycle-weight of $S$ is larger than $8c^*$. 
For a face $F$ of $H$, let $\lv(F)$ be the minimum level among the levels of the three vertices incident to $F$.
For a real number $\ell$ in range $[\ell_\textsf{min}, \ell_\textsf{max}-c^*]$, let $U_\ell$ be the union of $\cl(F)$ for all faces
$F$ with $\lv(F)<\ell$. 
Then $U_\ell$ does not contain $u$ because the level of any face incident to $u$ is at least $\ell_\textsf{max}-c^*$. 
We consider the connected component $F$ of $\mathbb R^2\setminus U_\ell$ containing $u$.
Let $C_\ell$ be the boundary curve of $F$.
See Figure~\ref{fig:simpcycc}(a--b).

\begin{lemma}\label{lem:cycle-level}
    For a vertex $w$ in $C_\ell$, 
	$\lv(\textsf{parent}(w)) < \ell \leq \lv(w)$. 
\end{lemma}
\begin{proof}
    If $\lv(w) < \ell$, 
    the levels of all faces $F$ incident to $w$ are lower than $\ell$. Then $F\subseteq U_\ell$ and $w\in \interior(U_\ell)$.
    This shows the second inequality.
    Let $F$ be a face incident to $w$ contained in $U_\ell$.
    Then $F$ is adjacent to a vertex $x$ with $\lv(x)<\ell$, and $(w,x)$ is an edge of $H$. 
    By the definition of level tree, $\lv(\textsf{parent}(w))\leq \lv(x)$.
    This shows the first inequality.
\end{proof}

\begin{lemma} \label{lem:cycle-like} 
The cycle $C_\ell$ is a simple cycle, and it intersects with $S$ at most twice. Moreover, if $\ell+2c^*<\ell_\textsf{max}$, then $C_\ell$ intersects with $S$ exactly twice.
\end{lemma}
\begin{proof} 
    Assume to the contrary that $C_\ell$ is not simple.
    By definition, $C_\ell$ is the boundary of a connected
    component of $\mathbb{R}_\ell\setminus U_\ell$. Therefore, the only possible case is that
    $U_\ell$ has a cut vertex $w$. 
    Then $F$ be a face of $H$ incident to $w$ lying outside of the component of $\cl(U_\ell)\setminus \{w\}$ containing $r$.
    See Figure~\ref{fig:simpcycc}(c).
    By the definition of $U_\ell$, there is a vertex $x$ incident to $F$ with $\lv(x)<\ell$.
    Consider the $r$-$x$ path in the level tree. The level of a vertex in the path is lower than $\ell$, and therefore, the path is contained in $\interior(U_\ell)$.
    Since this path must encounter $w$, and $\ell\leq \lv(w)$, $w$ is not a cut vertex of $C_\ell$.
    
    Next we show that $C_\ell$ intersects $S$ at most twice. Note that $C_\ell$ never contains $r$ by Lemma~\ref{lem:cycle-level}.
    If $|C_\ell \cap S|\geq 3$, we may assume that $|C_\ell \cap \pi_u|\geq 2$.
    Let $x,y$ be two vertices of $C_\ell\cap\pi_u$ with $\lv(x)>\lv(y)$. Then $y$ is an ancestor of $x$ in the level tree.
    Since both $x$ and $y$ does not lie on $\interior(U_\ell)$, their cycle-weights are at least $\ell$. 
    Then $\ell \leq\lv(y)\leq \lv(\textsf{parent}(x)) < \ell \leq \lv(x)$ by Lemma~\ref{lem:cycle-level}, which leads to a contradiction. 
    Finally, consider the case that $\ell+2c^* <\ell_\textsf{max}$.     
    The maximum level of vertices in $C_\ell$ is at most $\ell+c^* < \ell_\textsf{max}-c^*$ since $c(v)\leq c^*$ for any vertex $v\in V(H)$. 
    Observe that  both $\lv(v)$ and $\lv(u)$ are at least $\ell_\textsf{max}-c^*$. More precisely,
    $\lv(u)=\ell_\textsf{max}$ and $\lv(v)\geq \lv(u)-c(v)\geq \ell_\textsf{max}-c^*$. 
    Therefore, both $\pi_u$ and $\pi_v$ intersect $C_\ell$ at least once. By combining the previous argument,
    we conclude that they intersect $C_\ell$ exactly once. 
\end{proof}

\medskip
We compute a sequence $\mathcal D$ of cycles of small cycle-weight as follows. 
We divide the interval $[\ell_\textsf{min},\ell_\textsf{max}-c^*]$ into $t=\lfloor \frac{\ell_\textsf{max}-\ell_\textsf{min}}{c^*}-1\rfloor$ intervals, $I_1,\ldots, I_t$, such that
all intervals but the last one, have length exactly $c^*$. 	
For every odd indices $j\leq t-2$, we pick the number $\ell_j \in I_j$ that minimizes the cycle-weight of $C_{\ell_j}$. 
In this way, we obtain a sequence $\mathcal D=\langle C_{\ell_1},C_{\ell_3},C_{\ell_5}...,C_{\ell_{t'}}\rangle$.
To see that $\mathcal D$ is not empty, we observe that 
the cycle-weight of $S$ is $\lv(u)+\lv(v)-c(r)\leq 2\cdot\lv(u)$. Then we have 
$8c^* \leq  2\cdot \lv(u) =  2\ell_\textsf{max}$, and $\ell_\textsf{min}=c(r)\leq c^*$. 
Therefore, $3\leq t$, and $\mathcal D$ is not empty. 
Since we pick indices $j$ from $j\leq t-2$, we have $\ell_j+2c^* < \ell_\textsf{max}$.
We summary these facts into the following:
\begin{observation}
    $\mathcal D$ is not empty, and each cycle of $\mathcal D$ intersects with $S$ exactly twice.
\end{observation}
We show that $\mathcal D$ consists of vertex-disjoint cycles with small cycle-weight.

\begin{lemma} 
	The cycles in $\mathcal D$ are vertex-disjoint. 
\end{lemma}
\begin{proof}
     Let $C_{\ell_j}$ and $C_{\ell_{j+2}}$ be two consecutive cycles of $\mathcal D$.
    By definition, $\ell_{j+2}-\ell_{j}$ is in range $[c^*, 3c^*]$.
    For a vertex $v\in C_{\ell_j}$,
    $\lv(\textsf{parent}(v))<\ell_j$ by Lemma~\ref{lem:cycle-level}.
    In other words, $\lv(v)<\ell_j+c(v)$.
    Then $v$ is contained in $\interior(U_{\ell_j+c(v)})$.
    Since $\ell_j+c(v) \leq \ell_{j+2}$, 
        $v$ is contained in the interior of $U_{\ell_{j+2}}$, and thus it does not lie on $C_{\ell_{j+2}}$.
\end{proof}

\begin{lemma} \label{lem:prop-cycleseq}
	Each cycle of $\mathcal D$ has weight at most $c^*$.
\end{lemma}
\begin{proof} 
   We show the stronger statement that the sum of cycle-weights of all cycles in $\mathcal D$ is at most $c^*$.
    Let $c(C_\ell)$ be the cycle-weight of $C_{\ell}$, which is the sum of the cycle-weights of all vertices of $C_{\ell}$.
    For every index $j$ less than $t$, we have  
	\[c(C_{\ell_j}) \leq \frac{1}{c^*}\int_{\ell\in I_j}c(C_\ell), \text{ and thus } \sum_{1\leq j \leq t}{c(C_{\ell_j})} \leq \frac{1}{c^*}\int_{\ell\in [\ell_\textsf{min},\ell_\textsf{max}-c^*]} c(C_\ell).\] 

    For a vertex $w\in C_\ell$, $\lv(w) \geq \ell$ and $\lv(\textsf{parent}(w)) < \ell$. 
    Since $\lv(\textsf{parent}(w)) + c(w) = \lv(w)$, $\lv(w)\in [\ell, \ell+c(w)]$.
    Then each $w$ contributes at most $(c(w))^2$ to the integral of $c(C_\ell)$.
    Thus, the sum of all $c(C_{\ell_j})$ for $j\leq t$ is at most $c^*$.
\end{proof}

We can compute $\mathcal D$ in linear time as follows. Starting from $\ell_\textsf{min}$, imagine that we increase $\ell$ continuously.
Then $c(C_\ell)$ is a step function which maps $\ell\in[\ell_\textsf{min},\ell_\textsf{max}]$ to $c(C_\ell)$. This function has $|H|$ different values.
Therefore, we can compute the weight of $C_\ell$ for all $\ell$ in $O(|H|)$ time in total by sweeping the range $[\ell_\textsf{min}, \ell_\textsf{max}]$.

\begin{figure}
	\centering
	\includegraphics[width=0.65\textwidth]{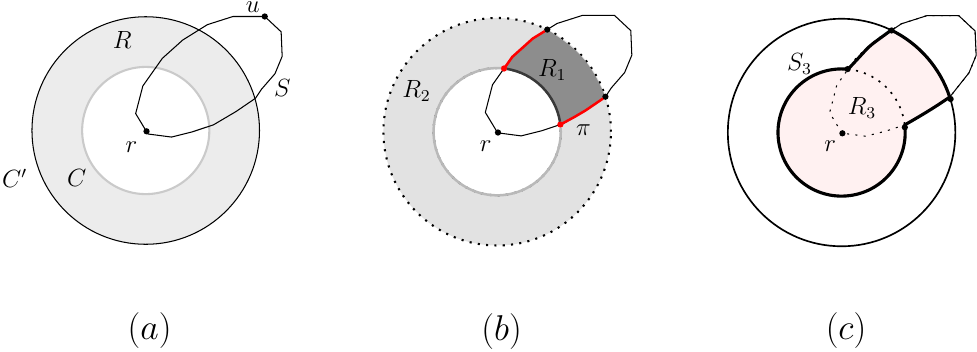}
	\caption{\small (a) A balanced cycle separator $S$ has cycle-weight larger than $8c^*$.
  The gray region $R$ is $C_2^r\setminus C_1^r$ for two consecutive cycles $C_1,C_2$ of $\mathcal D$.
		(b) Two regions $R_1$ and $R_2$, and the red paths $\pi=S\cap R$ partition $R$.
		(c) If $b(\pi)<\frac{1}{9}b(H)$ and $b(R_1)\geq b(R_2)$, $S_3$ is a $8/9$-balanced cycle separator.}
	\label{fig:cyclesep}
\end{figure}

\subsection{Balanced Cycle Separator with Small Cycle-Weight}
In this subsection, we compute a simple balanced cycle separator of small \emph{cycle-weight} using $\mathcal D$ similar to 
the algorithm of Har-Peled and Nayyeri~\cite{harpeled17cycleseparator}. 
For a subset $A$ of $\mathbb R^2$, we use $b(A)$ to denote the total balance-weight of the vertices of $H$ contained in $A$. 
By a slight abuse of notation, for a subgraph $H'$ of $H$, we let $b(H')$  be the total balance-weight of the vertices of $H'$.

Suppose there is a vertex $w$ with $b(w)\geq b(H)/9$. 
Then we consider a face $F$ incident to $w$. Three vertices incident to $F$ forms a desired separator because the sum of cycle-weights of three vertices is at most $3c^*$.
From now on, we assume that all vertices $w$ have the balance-weights $b(w)$ with $b(w)< b(H)/9$. 
We insert two cycles into $\mathcal D$: a trivial cycle consisting of the root $r$ of $\lt(H)$ in the front, and a trivial cycle consisting of $u$ of $\lt(H)$ in the last.
For a cycle $C$ in $\mathcal D$, no region of $\mathbb R^2\setminus C$ contains both $u$ and $r$.
Let $C^u$ and $C^r$ be two regions of $\mathbb R^2\setminus C$ contain $u$ and $r$, respectively.
If such a region $C^r$ (and $C^u$) does not exists, we set $C^r$ (and $C^u$) as the empty set.

If there is a cycle in $\mathcal D$ that is a $2/3$-balanced separator, we are done due to the Lemma~\ref{lem:prop-cycleseq}. 
Otherwise, we find two consecutive cycles $C_1$ and $C_2$ in $\mathcal D$ such that $b(C_1^r) < \frac{1}{3}b(H)$ and $b(C_2^r)>\frac{2}{3}b(H)$. This is always possible because two trivial cycles have balance weight at most $\frac{1}{9}b(H)$.
Let $S$ be the cycle specified in Lemma~\ref{lem:largecyclesep}.
We also specify some subsets and paths as follows.
Let $R=C_2^r \setminus C_1^r$, $R_1=R\cap S^\textsf{in}$, $R_2=R\cap S^\textsf{ex}$, $R_3=R_1\cup C_1^r$, and $R_4=R_2\cup C_1^r$.
See Figure~\ref{fig:cyclesep}. Let $\pi=S\cap R$ be the subset of the cycle-separator $S$.
Since $C_1$ and $C_2$ are consecutive cycles in $\mathcal D$, the total cycle-weights of the vertices of $\pi$ is at most $4c^*$. 
The boundary $\partial R_i$ consists of a single cycle, namely $S_i$. Then the total cycle-weights of the vertices in $S_i$ is at most $10c^*=O(c^*)$ due to Lemma~\ref{lem:prop-cycleseq}.

We claim that one of $S_i$'s is a $8/9$-balanced cycle separator of $H$.
Since $S$ is a $2/3$-balanced separator, $b(R_1)$, $b(R_2)$ is at most $\frac{2}{3}b(H)$.
For the first case that $b(R_i)\geq \frac{1}{3}b(H)$ for $i\in \{1,2\}$, $S_i$ is a $2/3$-balanced cycle separator.
For the second case that total balance-weights of vertices in $\pi$ is at least $\frac{1}{9}b(H)$, $S_1$ is a $8/9$-balanced cycle separator because all vertices of $\pi$ are contained in $S_1$.
The remaining case is that $b(R_1),b(R_2)<\frac{1}{3}b(H)$ and $b(\pi)<\frac{1}{9}b(H)$.
Since $\{R_1, \pi, R_2\}$ forms a partition of $R$, we have 
\[b(R_1)+b(R_2)=b(R)\setminus b(\pi)\geq \frac{1}{3}b(H)-\frac{1}{9}b(H)=\frac{2}{9}b(H). \] 
Then we can pick $i\in \{1,2\}$ so that $b(R_i)\geq \frac{1}{9}b(H)$.
Then 
\[ 
\frac{1}{9}b(H) \leq b(R_i) < b(R_{i+2})=b(R_i)+b(C_1^r)<\frac{2}{3}b(H). 
\]
Therefore, $S_{i+2}$ is a desired separator.

\begin{lemma} \label{lem:cycle-sep}
    For a triangulated plane graph $H$ with cycle-weight $c(\cdot)$ and balance-weight $b(\cdot)$ satisfying
    $1\leq c(v)\leq |H|^{O(1)}$ and $0\leq b(v)$,
    we can compute a $(8/9)$-balanced cycle separator of weight $10\cdot \sqrt{\sum_{v\in V(H)}(c(v))^2}$ in $O(|H|)$ time. 
\end{lemma}

\section{Surface Decomposition of Small Weighed Width} \label{sec:annulus}
In this section, we show that a plane graph $H$ with cycle-weights $c: V(H)\rightarrow \mathbb{R}$ with $1\leq c(v)\leq |H|^{O(1)}$  admits a surface decomposition 
of weighted width $O(\wsqr{H}+\wmax{H}^4)$. 
For most of applications, the first term in the weighted width dominates the second term. 
Recall that a surface decomposition of $H$ is a recursive decomposition of $\mathbb{R}^2$ into pieces such that all pieces in the lowest level have at most two vertices of $G$, and the boundary of no piece is crossed by an edge of $H$.
We represent it as a pair $(T, \mathcal A)$ where $T$ is a rooted binary tree, and $\mathcal A$ is a mapping that maps a node $t$ of $T$ into a piece $A_t$. 
 The \emph{weight} of a node $t$ is defined as the sum of weights of the vertices in $V(H)\cap \partial A_t$. 
    The \emph{weighted width} of $(T,\mathcal A)$ is defined as the maximum weight of the nodes of $T$.

We first give a sketch of our algorithm. 
Initially, we have the tree consisting of a single node $t$ associated with $A_t=\mathbb R^2$ and $H_t=H$. 
At each iteration, for a leaf node $t$, we compute a balanced simple cycle separator $S_t$ using Lemma~\ref{lem:cycle-sep} 
by assigning balance-weights to the vertices of $H_{t}$ properly. 
Let $\gamma_t$ be a Jordan curve intersecting $H_t$ only at vertices of $S_t$. 
Then 
$A_t$ is partitioned into connected components by $\gamma_t$. 
We add a child $t'$ of $t$ each corresponding to each connected component.
That is, we let  $A_{t'}$ be the closure of a connected component of $A_{t}-\gamma_{t}$,
and $H_{t'}$ be the subgraph of $H_{t}$ induced by $A_{t'}\cap V(H)$. 
We do this until $H_t$ has at most $\max\{(\wmax{H})^3,10^{24}\}$ vertices. When we reach a node $t$ such that $H_t$ has at most $\max\{(\wmax{H})^3,10^{24}\}$ vertices, we can compute the descendants of $t$ in a straightforward way such that the weight of the resulting surface decomposition increases by $|V(H_t)|{\wmax{H_t}} \leq O({(\wmax{H_t})^{4}})$.

There are three issues: First, $H$ is not necessarily triangulated, so we cannot directly apply Lemma~\ref{lem:cycle-sep}. 
Second, although a single balanced cycle separator of  Lemma~\ref{lem:cycle-sep} has complexity $\wsqr{H_t}$, $H_{t}$ 
might have $\Theta(d\cdot \wsqr{H_t})$ vertices lying on the boundary of $A_t$, and $A_t$ might have rank $d$ in the worst case for a node $t$ constructed in the $d$th recursion step for $d=\omega(1)$. For the former case, the weighted width of the surface decomposition computed by this approach can exceed the desired value, and for the latter case, $A_t$ violates the conditions for being a piece. 
Third, the number of components of $A_t-\gamma_t$ can be more than two. In this case, the previous approach creates more than two children, which violates the conditions for the surface decomposition.

The first issue can be handled by triangulating $H$ and assigning weights carefully to the new vertices. 
The second issue can be handled by assigning balance-weights so that, in each recursion step, 
the complexity of $S_t$ decreases by a constant factor, or the rank of $A_t$ decreases by a constant factor. 
Then the third issue can be handled by partitioning the components into exactly two groups, and then adding two children for the two groups. 
We introduce the tools for dealing with the first two issues in Section~\ref{sec:holesep}
and show how to handle all the issues to compute a surface decomposition of desired weighted width in Section~\ref{sec:annulus-construction}.

\subsection{Vertex and Hole Separators for a Piece} \label{sec:holesep}
Let $H_t$ be the subgraph of $H$ induced by $A_t\cap V(H)$ with $|V(H_t)|\geq \max\{(\wmax{H_t})^3, 10^{24}\}$. 
In this subsection, we introduce
two types of separators of $H_t$ called a \emph{vertex separator} and a \emph{hole separator}. 
A \emph{vertex separator} $\gamma_\textsf{v}$ is a noose of $H_t$ such that the total cycle-weight of the vertices contained in the \emph{closure} of each connected component of $\interior(A_t)-\gamma_\textsf{v}$ is at most $9/10$ of the total cycle-weight of $H_t$. 
A \emph{hole separator} $\gamma_\textsf{h}$ is a noose of $H_t$ 
such that the rank of the closure of each connected component of $\interior(A_t)-\gamma_\textsf{h}$ is at most $9/10$ of the rank of $A_t$. This is well-defined due to the following lemma. 
We can compute them by setting balanced-weights $b(\cdot)$ of the vertices of $H_t$ carefully and then
using the balanced cycle separator described in Section~\ref{sec:cycle-separator}.

\begin{lemma}\label{lem:jordan-annulus}
    For a closed curve $\gamma$ intersecting a piece $A$, 
    the closure of each connected component of $\interior(A)-\gamma$ is also a piece with rank at most one plus the rank of $A$. Moreover, the union of the closures of the connected components of $\interior(A)-\gamma$ is $A$ itself. 
\end{lemma}
\begin{proof}
    For the first claim, let $F$ be a connected component of $\interior(A)-\gamma$. 
    Since $F$ is an open set, $\cl(F)$ is interior-connected.     
    Also, since $F$ is a connected open point set in the plane, $\interior(\cl(F))=F$, and thus $\cl(F)$ is regular closed. 
    Finally, the rank of $\cl(F)$ is at most one plus the rank of $A$. 
    To see this, we call a connected component of $\mathbb{R}^2\setminus A$ a \emph{hole} of $A$. 
    The rank of $A$ is the number of holes of $A$ by definition. 
    A hole of $A$ is contained in a hole of $F$.
    Also, one of the regions cut by $\gamma$, say $X$, is contained in a hole of $F$.
    On the other hand, each hole of $F$ intersects either a hole of $A$ or $X$. Therefore, 
    $\cl(F)$ is a piece, and its rank is at most one plus the rank of $A$. 

    For the second claim, let $F_1,\ldots, F_\ell$ be the connected components of $\interior(A)-\gamma$. Clearly, $\cl(F_1)\cup\ldots \cup \cl(F_\ell) \subseteq A$ because $F_1,\ldots, F_\ell \subseteq A$ and $A$ is closed. 
    For the other direction, let $p$ be a point of $A$.
    If $p$ lies on $\gamma$, there is an index $i$ such that $F_i$ is incident to $\gamma$ at $p$, and thus $p$ is contained in $\cl(F_i)$.
    If $p$ is contained in $\interior(A)-\gamma$, it is contained in $F_1\cup\ldots \cup F_\ell$, and thus it is contained in $\cl(F_1)\cup\ldots \cup \cl(F_\ell)$. 
    If $p$ is contained in $\partial A-\gamma$, there is a small disk centered at $p$ intersecting $\interior(A)-\gamma$ at a point, say $q$. Then
    there is an index $i$ such that $F_i$ contains $q$, and moreover, $\cl(F_i)$ contains $p$.
    For any case, $p$ is contained in $\cl(F_1)\cup\ldots \cup \cl(F_\ell)$, and thus the second claim also holds. 
\end{proof}

  \begin{figure} 
	\centering
	\includegraphics[width=0.7\textwidth]{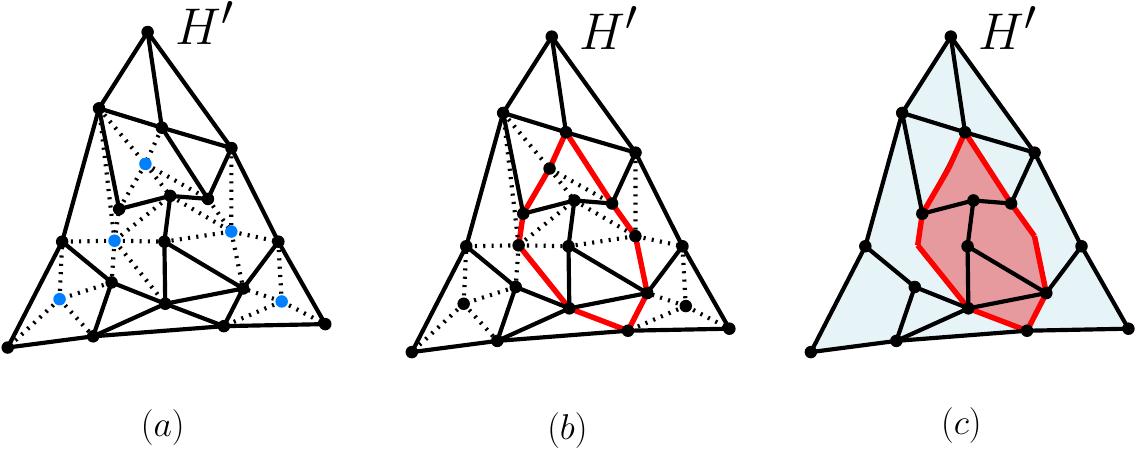}
	\caption{\small (a) Illustrates the construction of $H_\textsf{tr}$. Auxiliary vertices are denoted by blue points.
		(b) Illustrates a balanced cycle separator (red cycle). (c) The balanced cycle separator subdivides the piece into two smaller pieces. 
	}
	\label{fig:cycle-like-separator}
\end{figure}

We first triangulate $H_t$ by adding  an auxiliary vertex $v'$ in the interior of each face of $H_t$, and connecting it with each vertex incident to the face by an edge.\footnote{Here, an edge is not necessarily drawn as a single  segment. It is not difficult to see that we can triangulate each face $F$ of $H_t$ by drawing the edges as polygonal curves of total complexity polynomial in the complexity of $F$.} Let $H_{\textsf{tr}}$ be the triangulation of $H_t$.
Since the auxiliary vertices
do not have cycle-weights yet, we set the cycle-weight $c(v')$ as one.
Note that 
$\wsum{H_\textsf{tr}}\leq 3\wsum{ H_\textsf t}$ and $\wsqr{H_\textsf{tr}}\leq \sqrt 3\wsqr{H_\textsf t}$,
because the number of faces of $H$ is at most $2|V(H)|$ due to Euler's formula.

\subparagraph{Vertex separator.} 
To obtain a vertex separator, we set $b(v)$ to $c(v)$ for every vertex $v$ of $H_{\textsf{tr}}$ which comes from $H_t$, and 
we set $b(v')$ to zero for every auxiliary vertex $v'$ of $H_\textsf{tr}$.
We compute a cycle separator $S$ of cycle-weight $10\cdot \wsqr{H_{\textsf{tr}}}\leq 10\sqrt 3\cdot \wsqr{H_t}$ of $H_\textsf{tr}$ using Lemma~\ref{lem:cycle-sep}. 
Let $S_\textsf{v}$ be the sequence of vertices of $S$ excluding the auxiliary vertices.
Then $S_\textsf{v}$ is a balanced separator of $H_t$ by construction. Moreover, we can draw a noose $\gamma_\textsf{v}$ of $H_t$ intersecting all vertices of $S_\textsf{v}$. 
See Figure~\ref{fig:cycle-like-separator}(c). 

Notice that $\gamma_\textsf{v}$ partitions $A_t$ into at least two regions.  
Let $H'$ be the subgraph of $H_t$ induced by the vertices in the \emph{closure} of a connected component of $\interior(A_t)-\gamma_\textsf{v}$. 
By Lemma~\ref{lem:cycle-sep}, the sum of $b(v)$ for all vertices in $H'$ 
not lying on $\gamma_\textsf{v}$ is at most $(8/9)\cdot \wsum{H_t}$ as the sum of $b(v)$ for all vertices $v\in H_{\textsf{tr}}$ is exactly $\wsum{H_t}$ by the choice of $b(\cdot)$. 
However, the vertices of $H'$ lying on $\gamma_\textsf{v}$ also have cycle-weight, but they are not considered in Lemma~\ref{lem:cycle-sep}. 
But the total cycle-weight of such vertices is sufficiently small. 
Specifically, the cycle-weight of $\gamma_\textsf v$ is at most  $10\sqrt 3\wsqr{H_t}\leq 10\sqrt 3\cdot \sqrt{\wmax{H_t}^2 \cdot |H_t|}= 10\sqrt 3|H_t|^{5/6}$ by the assumption. 
Moreover, we have $|H_t|^{5/6} \leq 10^{-4} \cdot |H_t| \leq 10^{-4} \cdot (\wsum{H_t})$ by the assumption that $c(\cdot)\geq 1$ and $|H_t|\geq 10^{24}$. 
Therefore, the total cycle-weight of $H'$ is at most $(9/10) \cdot \wsum{H_t}$, and thus 
 $\gamma_\textsf{v}$ is a vertex separator of $H_t$. 

\subparagraph{Hole separator.}
To obtain a hole separator, we set $b(v)=0$ for every vertex $v$ of $H_\textsf{tr}$ which comes from $H_t$.
A face of $H_t$ is either contained in $A_t$, or contains a connected component of $\mathbb{R}^2-A_t$. 
We set $b(v')=0$ for a vertex $v'$ corresponding to a face of $H_t$ of the former type, and 
set $b(v')=1$ for a vertex $v'$ corresponding to a face of $H_t$ of the latter type. 
Since the number of faces of the latter type is exactly the rank of $A_t$, the total balance-weight of $H_{\textsf{tr}}$ is equal to the rank of $A_t$.
Then we compute a balanced cycle separator of weight $O(\wsqr{H_{\textsf{tr}}})$ of $H_\textsf{tr}$. 
As we did for a vertex separator, we can obtain a balanced separator $S_\textsf{h}$ of $H_t$
and a noose $\gamma_\textsf{h}$ intersecting the vertices of $S_\textsf{h}$. 
By construction, the rank of the closure of each connected region in $\interior(A_t)-\gamma_\textsf{h}$ is at most $8/9$ of the rank of $A_t$ plus one, and thus $\gamma_\textsf{h}$ is a hole separator.

\begin{lemma} \label{lem:annulus-separator}
	Let $H$ be a planar graph with cycle-weight $c$  drawn in a piece 
        with $|H|\geq 10^{24}$  and $1\leq c(v) \leq {|H|}^{1/3}$ for all vertices $v\in V(H)$.
        Then we can compute a vertex separator and a hole separator of 
	weight $O(\sqrt{\sum_{v\in V(H)}(c(v))^2})$ in $O(|H|)$ time.
\end{lemma}

\subparagraph{Remark.} In our application of vertex and hole separators in Section~\ref{sec:annuluscarving},
there might be \emph{forbidden polygonal curves}. A forbidden polygonal curve is a polygonal curve which is not a part of the drawing of $H$, but intersects $H$ only at a single vertex of $H$. 
It is not difficult to see that we can construct 
a vertex and hole separator not intersecting any forbidden polygonal curves (except for their intersection with $V(H)$) without increasing the weight of the separators. 

\subsection{Recursive Construction} \label{sec:annulus-construction}
\begin{figure}
		\centering
		\includegraphics[width=0.7\textwidth]{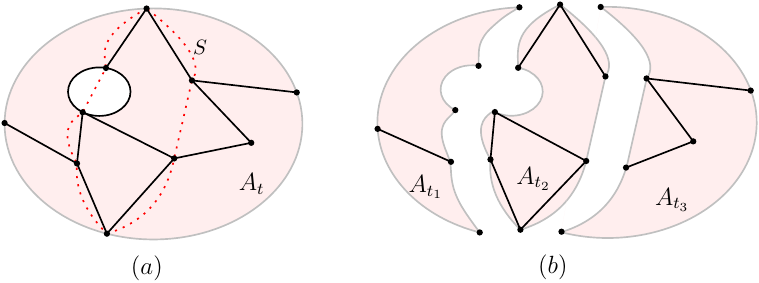}
		\caption{\small 
		(a) The boundary of the red piece $A_t$ intersects $H$ only at the vertices. The red dotted curve denotes the separator $S$.
		(b) First we separate $A_{t_1}$ and $(A_{t_2}\cup A_{t_3})$, and then again separate $A_{t_2}$ and $A_{t_3}$. The boundaries of pieces again intersect $H$ only at the vertices.
        }
		\label{fig:annulusdecomp}
	\end{figure}

In this subsection, we give an $O(|H|\log |H|)$-time algorithm that computes a surface decomposition $(T, \mathcal A)$ of $H$ of weighted width $O(\wsqr{H}+(\wmax{H})^4)$. 
As mentioned before, the algorithm starts from the tree $T$ of a single node $t$ with $A_t=\mathbb R^2$ and $H_t=H$.
For each leaf node $t$, we subdivide $A_t$ further if $A_t$ contains more than one edge of $H$. 
If $H_t$ has more than $\max\{(c_\text{max}(H))^4,10^{24}\}$ vertices, we do the following.
If the rank of $A_t$ exceeds some constant, say $100$, we compute a hole separator $\gamma$ of $H_t$. 
Otherwise, we compute a vertex separator $\gamma$ of $H_t$. 
We denote the set of the closures of the connected components of $\interior(A_t)-\gamma$
by $\mathcal A_\gamma$.  
See Figure~\ref{fig:annulusdecomp}(a). Each region of $\mathcal A_\gamma$ is a piece as shown in Lemma~\ref{lem:jordan-annulus}, but the size of $\mathcal A_\gamma$ can be more than two.  
One might want to add $|\mathcal A_\gamma|$ children of $t$ and assign each piece of $\mathcal A_\gamma$ to each child. However, each internal node of $T$ for a surface decomposition $(T,\mathcal A)$ should have exactly two children by definition.
Thus instead of creating $|\mathcal A_\gamma|$ children of $t$, we construct a binary tree rooted at $t$ 
whose leaf edges correspond to the pieces of $\mathcal A_\gamma$. 
Here, since a piece must be regular closed and interior-connected, we have to construct such a binary tree carefully.

For this purpose, consider the adjacency graph of $\mathcal A_\gamma$ where the vertices represents the pieces of $\mathcal A_\gamma$ and two vertices are connected by an edge if their respective pieces are adjacent. See Figure~\ref{fig:annulusdecomp}(b). 
Every connected graph has a vertex whose removal does not disconnect the graph: choose a leaf node of a DFS tree of the graph. We compute a DFS tree of the adjacency graph and choose a leaf node of the tree. Let $A$ be the piece corresponding to the chosen leaf node. 
Then $A$ and the union $A'$ of the pieces of $\mathcal A_\gamma-\{A\}$ are also pieces.
We create two children of $t$, say $t_1$ and $t_2$, and then let $A_{t_1}=A$ and $A_{t_2}=A'$, respectively.  
Then we update the DFS tree by removing the node corresponding to $A$, and choose a leaf node again. We add two children of $t_2$ and assign the pieces accordingly.
We do this repeatedly until we have a binary tree of $t$ whose leaf nodes have the pieces of $\mathcal A_\gamma$. 
After computing the binary tree, we let $H_{t'}=H_t[V(H_t)\cap A_{t'}]$ for each node $t'$ of the binary tree. 

If the number of vertices of $H_t$ is at most $\max\{(\wmax{H})^3,10^{24}\}$, we can compute the descendants of $t$ in a straightforward way so that the weight of the surface decomposition increases by $(\wmax{H})^4+O(1)$, and maintain the rank of the piece as a constant. 
In particular, we compute an arbitrary noose that divides the piece into two pieces such that the number of vertices of $V(H_t)$ decreases by at least one.
Whenever the rank of the piece exceeds a certain constant, we compute a hole separator that reduces the rank of the piece by a constant fraction, and then compute the descendants as we did for the case that $H_t$ has more than  $\max\{(\wmax{H})^3,10^{24}\}$ vertices. 
By repeating this until all pieces in the leaf nodes have exactly one edge of $H$, we can obtain $(T,\mathcal A)$ satisfying the conditions \textbf{(A1--A3)} for being a surface decomposition of $H$. Thus in the following, we show that the weighted width of   $(T,\mathcal A)$ is $O(\wsqr{H}+(\wmax{H})^4)$, and the surface decomposition can be computed in $O(|H|\log |H|)$ time. 


\begin{lemma} \label{lem:annulus_time}
    We can compute $(T, \mathcal A)$ in $O(|H|\log |H|)$ time if $\wmax{H} \leq |H|^{O(1)}$.
\end{lemma}
\begin{proof}	
    Consider the recursion tree of our recursive algorithm. For each recursion step for handling a node $t$, 
    we first compute a separator of $H_t$ in $O(|H_t|)$ time.
    Then the separator splits $A_t$ into smaller pieces, and we compute a DFS tree of the adjacency graph of the smaller pieces in $O(|H_t|)$ time. 
    For all recursive calls in the same recursion level, the total complexity of $H_t$'s is
    $O(|H|)$. Therefore, the recursive calls in one recursion level take $O(|H|)$ time.

    Then we claim that the recursion depth is $O(\log |H|)$, which leads to the total time complexity of $O(|H|\log |H|)$. 
    Consider a longest path in the recursion tree. For each recursive call in this path,
    we compute either a vertex separator or a hole separator: If the rank of $A_t$ is larger than 100, we compute a hole separator, and otherwise, we compute a vertex separator. 
    If we compute a vertex separator at some level, the total cycle-weights of $H_{t'}$ in the next recursive call (for a node $t'$ of $T$) in this path decreases by a constant factor.
    Note that once we use a hole separator at some level, the rank of $A_{t'}$ decreases to at most $\frac{8}{9}\cdot 100+1\leq 90$, and thus no two consecutive calls in the path use both hole separators. That is, the total cycle-weight of $H_t$ decreases by a constant factor for every second recursive call in the path. Therefore, the height of the recursion tree
    is $O(\log \wmax{H})=O(\log |H|)$. 
\end{proof}

\begin{lemma} \label{lem:algo-annuluswidth}
	$(T, \mathcal A)$ has weighted width  $O(\sqrt{\sum_{v\in V(H)}(c(v))^2})$.
\end{lemma}
\begin{proof}
    Recall that the weight of a node $t$ of $T$ is the sum of the cycle-weights of $\bdr t$, where $\bdr t$ is the set of vertices of $H_t$ contained in $\partial A_t$. 
    We again consider the recursion tree of our recursive algorithm. 
    By construction, $\partial A_t$ is contained in the union
    of the hole and vertex separators constructed for the ancestors of the recursion tree of the recursive call that constructs $t$. 
    As shown in the second paragraph of the proof of Lemma~\ref{lem:annulus_time},
    the total-cycle weight of $H_t$ decreases by a constant fraction for every second recursive call in a path in the recursion tree. 
    That is, the cycle-weight of $H_t$ is (almost) geometrically decreasing along the path.
    Therefore, the total weight of the hole and vertex separators constructed for the ancestors is bounded by the weight of the hole or vertex separator constructed for the root of the recursion tree, which is $O(\sqrt{\sum_{v\in V(H)}(c(v))^2})$.
\end{proof}

Theorem~\ref{thm:annulus} summarizes this section.
\annulus*

 \section{Surface Cut Decomposition of a Graph with ICF-Property}\label{sec:annuluscarving}
    Let $G=(V,E)$ be a graph given with its straight-line drawing in the plane that admits the \textbf{icf}-property. 
    In this section, 
    we present a polynomial-time algorithm that computes an sc-decomposition of clique-weighted width $O(\sqrt{\ell})$ with respect to $\mathcal M$ assuming that
    a map $\mathcal M$ of $G$ is given, where $\ell$ is 
    the number of vertices of degree at least three in $G$.
    Let $H$ be a map sparsifier of $G$ with respect to $\mathcal M$.

    A key idea is to use a surface decomposition $(T, {\mathcal A})$ of $H$. 
    For each piece $A_t$ of ${\mathcal A}$, the boundary curves of $A_t$ intersect $H$ only at vertices of $H$. 
    We slightly perturb the boundary curves locally so that no piece contains a vertex of $G$
    on its boundary. 
    Then we can show that the resulting pair, say $(T,  \bar{\mathcal A})$, is an sc-decomposition of $G$.
    By defining the weights of the vertices of $H$ carefully, we can show that the clique-weighted width of the sc-decomposition of $G$
    is at most the weight of $(T, {\mathcal A})$. 
    However, the sum of the squared weights of the vertices of ${H}$ can be $\Theta(n)$ in the worst case, and thus the weight of $(T, {\mathcal A})$ is $O(\sqrt n)$. 
    To get an sc-decomposition of $H$ of weight $O(\sqrt{\ell})$, we consider the minor $H_3$ of $H$ obtained by 
    contracting each maximal path consisting of degree-1 and degree-2 vertices and then by removing all degree-1 vertices in the
    resulting graph. 
    Then we apply the algorithm in Section~\ref{sec:annulus} to $H_3$ to obtain a surface decomposition of $H_3$ of weight $O(\sqrt{\ell})$. Then using it, we reconstruct a surface decomposition of $H$ of weight $O(\sqrt{\ell})$ as mentioned earlier. 

    Although this basic idea is simple, there are several technical issues to implement the idea and prove the correctness of our algorithm.
    In Section~\ref{sec:step-1}, we define the weight of the vertices of $H$ and construct the minor $H_3$ of $H$. 
    We analyze the sum of the squared weights of the vertices of $H_3$, and this gives an upper bound on the surface decomposition of $H_3$ constructed from Section~\ref{sec:annulus}. 
    In Section~\ref{sec:step-2}, we construct a surface decomposition $(T, {\mathcal A})$ of $H$ using the surface  decomposition of $H_3$, and analyze its width. In Section~\ref{sec:step-3}, we modify $(T, {\mathcal A})$  to construct an sc-decomposition $(T, \bar{\mathcal A})$ of $G$ by perturbing the boundary curves of the pieces of $\mathcal A$.

     

    \subsection{Step 1: Construction of \texorpdfstring{$H_3$}{Lg}} \label{sec:step-1}
    We first define the weight of the vertices of $H$ so that 
    the clique-weighted width of the sc-decomposition of $G$ constructed from a surface decomposition $(T, {\mathcal A})$ of $H$ is at most the weight of $(T, {\mathcal A})$.     
    For each cell $M$ of $\mathcal M$, we define the weight $c_H(v)$ of a vertex $v$ contained in $M$
    as
    the sum of the clique-weights of all $\alpha$-neighboring cells of $M$ in $\mathcal M$. 
    That is, $c_H(v)=1+\sum\log(1+|V(G)\cap M'|)$ where in the summation, we take all $\alpha$-neighboring cells $M'$ of $M$ in $\mathcal M$. 
    Recall that a cell $M'$ is an $r'$-neighboring cell of $M$ if a shortest path between $M$ and $M'$ has length at most $r'$ in the dual graph of $\mathcal M$. By the definition of a map, $M$ has $O_{r'}(1)$ $r'$-neighboring cells in $\mathcal M$. 
    

    To compute $H_3$ from $H$, we compute all maximal paths of $H$ whose internal vertices have degree exactly two.
    Among them, we remove all paths one of whose endpoints has degree one in $H$ (excluding the other endpoint having degree larger than two).
    For the other paths, we contract them into single edges (remove all internal vertices and connect the two endpoints by an edge.)
    The resulting graph is denoted by $H_3$.
    Note that $H_3$ might have a vertex of degree less than three, but such a vertex has degree at least three in $H$. 
    To apply
    the notion of the surface decomposition, we draw $H_3$ in the plane as in the drawing of $H$: the drawing of $H_3$ is a subdrawing of the drawing of $H$. Some points in the drawing considered as vertices in $H$ are not considered as vertices of $H_3$, and some polygonal curve in the drawing
    of $H$ is removed in $H_3$ if its endpoint has degree one in $H$. 
    We compute a surface decomposition $(T,  {\mathcal A})$ of $H_3$ using Theorem~\ref{thm:annulus}. The decomposition has width $O(\sqrt \ell)$ by Lemma~\ref{lem:sum-cost}.

    \begin{restatable}{lemma}{sumcost}
    \label{lem:sum-cost}
        We have 
        {$\wmax{H_3 } \leq \wmax{H} = O(\log \ell)$} and 
        $\wsqr{H_3} = O(\sqrt{\ell})$.
    \end{restatable}
    \begin{proof}
    Let $v$ be a vertex of $H$ contained in a cell $M_v$ of $\mathcal M$. Its weight $c_H(v)$ is the total clique-weight of $O(1)$-neighboring cells of $M_v$ in $\mathcal M$. 
    The number of such cells is $O(1)$ due to the definition of the map. 
    For the first statement, since $V(H_3)$ is a subset of $V(H)$, the inequality holds immediately. 
    For an $O(1)$-neighboring cell $M$ of $M_v$,
    if $|M\cap V(G)|\leq 3$, it contributes $O(1)$ to $c_H(v)$.
    If $|M\cap V(G)| > 3$, all vertices of $G$ contained in $M$ have degree at least three in $G$. Therefore, such a cell contributes $\log (1+ |V(G)\cap M|)=O(\log \ell)$ to $c_H(v)$. 
    Therefore, $c_H(v)=O(\log \ell)$, and thus $\wmax{H} = O(\log \ell)$.
    
    For the second statement,     
    observe 
    the number of vertices of $H_3$ is $O(\ell)$ by Lemma~\ref{lem:map-spanner-complexity} since all vertices of $H_3$ have degree at least three in $H$. 
    Also, we have $c(v)^2 = (1+ \sum \log (1+|V(G)\cap M|))^2 \leq (1+\sum (\log(1+|V(G)\cap M|))^2)\cdot O(1)$ by the Cauchy-Schwarz inequality.
    Therefore, each cell $M$ contributes $O(\log (1+|V(G)\cap M|)^2)$ to $c(v)^2$.    
    Each cell is an $O(1)$-neighboring cell of $O(1)$ cells, and each cell contains $O(1)$ vertices of $H$ (and thus $H_3$).
    Thus $M$ contributes to the weights of $O(1)$ vertices of $H_3$. 

    Consider the cells in $\mathcal M$ containing at most three vertices of $G$, say \emph{light cells}. 
    Each light cell contributes $O(1)$ to $c(v)^2$ since $\log (1+|V(G)\cap M|) \leq \log 4=2$. 
    Thus, for each vertex $v$ of $H_3$, the portion of $c(v)^2$ induced by the light cells is $O(1)$, and  the portion of the sum of $c(v)^2$ induced by light cells is $O(\ell)$ over all vertices of $H_3$ in total. 
    Now consider the cells in $\mathcal M$ containing more than three vertices of $G$, say \emph{heavy cells}.
    Since $(\log z+1)^2\leq 2z$ for every natural number $z$, 
    a heavy cell $M$ contributes at most $2|V(G)\cap M|$ to a vertex of $H_3$ contained in an $O(1)$-neighboring cell of $M$. 
    Notice that the sum of $2|V(G)\cap M|$ over all heavy cells $M$ is $O(\ell)$ since the number of vertices of $H_3$ is $O(\ell)$. 
    Therefore, the heavy cells contribute $O(\ell)$ to the sum of $c(v)^2$ over all vertices $v$ of $H_3$ in total.   
    This implies that the sum of $c(v)^2$ over all vertices $v$ of $H_3$ is $O(\ell)$, and $\wsqr{H_3}$, which is the 
    square of the sum of $c(v)^2$, is at most $O(\sqrt \ell)$.
    \end{proof}

    By construction of the drawing of $H_3$, some polygonal curves of the drawing of $H$ do not appear in the drawing of $H_3$. Thus the boundary of a piece of ${\mathcal A}$ might cross an edge of $H$.
    We can avoid this case by constructing a surface decomposition of $H_3$ more carefully. A polygonal curve of $H$ not appearing in $H_3$ is attached at a single vertex of $H_3$. Whenever we construct a hole and vertex separator during the construction of a surface decomposition of $H_3$, we set those polygonal curves as forbidden regions. We can construct a hole and vertex separator not intersecting these forbidden regions as mentioned in the remark at the end of Section~\ref{sec:holesep}.

    \subsection{Step 2: Constructing a Surface Decomposition of \texorpdfstring{$H$}{Lg} Using \texorpdfstring{$H_3$}{Lg}}\label{sec:step-2}
\begin{figure}
		\centering
		\includegraphics[width=0.7\textwidth]{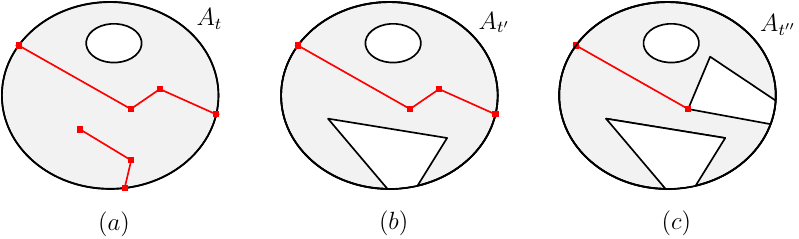}
		\caption{\small (a) The piece $A_t$ contains a traversing chain and an attached chain. 
        (b) We first handle the attached chain. The new closed curve forms a part of the outer boundary curve of $A_{t'}$. 
        (c) Then we handle the traversing chain. We compute a curve separating an edge of the traversing chain from the other edges. 
        Since $A_{t''}$ contains exactly one edge of $H$, we do not subdivide $A_{t''}$ further. 
		}
		\label{fig:BaseNode}
    \end{figure}

    Given a surface decomposition ($T$,  ${\mathcal A}$) of $H_3$ intersecting the drawing of $H$ only at its vertices, 
    we construct a surface  decomposition of $H$. We simply call a maximal path of $H$ whose internal vertices have degree two 
    a \emph{maximal chain} of $H$.  If both endpoints of a maximal chain lie on the boundary of $A_t$ for a node $t$, it is called 
    a \emph{traversing chain} of $A_t$. Otherwise, exactly one endpoint lies on the boundary of $A_t$. In this case, it is called an \emph{attached chain} of $A_t$. See Figure~\ref{fig:BaseNode}(a). 
    A piece of a leaf node of $T$ might contain maximal chains of $H$. 
    Recall that by the definition of a surface decomposition, each leaf node must have a piece containing at most two vertices of $H$. 
    To handle this issue, we subdivide such pieces further.
    Once this is done, the resulting pair becomes a surface decomposition of $H$. 

    Let $t$ be a leaf node of $T$ such that $A_t$ contains a maximal chain.
    Here, $A_t$ contains at most one traversing chain, but 
    it might contain more than one attached chains. This is because the endpoints of a traversing chain 
    appear in $V(H_3)\cap A_t$.
    We handle each maximal chain one by one starting from the attached chains (if they exist). 
    To handle a maximal chain $\pi$, we cut $A_t$ into two pieces along a closed polygonal curve $C$ whose interior 
    is contained in $A_t$ as follows. 
    If there is a maximal chain in $A_t$ other than $\pi$, we separate $\pi$ from all other maximal chains. 
    Note that $\pi$ is an attached chain by definition. 
    In this case, we choose $C$ such that it intersects $\pi$ only at the endpoint $p$ of $\pi$ lying on the boundary of $A_t$, 
    and it touches $\partial A_t$ along an $\epsilon$-neighborhood of $p$ in $\partial A_t$ for a sufficiently small constant $\epsilon>0$ so that the rank of the piece does not increases. 
    See Figure~\ref{fig:BaseNode}(b). 
    If $\pi$ is a unique maximal chain in $A_t$, we separate one endpoint of $\pi$ from all other edges of $\pi$. 
    We choose $C$ such that $C$ intersects $\pi$ only at the first vertex of $\pi$ and the second last vertex of $\pi$, and it
    touches $\partial A_t$ along an $\epsilon$-neighborhood of the first vertex of $\pi$ in $\partial A_t$ for a sufficiently small
    constant $\epsilon>0$. Here, $\pi$ is oriented from an arbitrary endpoint lying on $\partial A_t$ to the other one.
    In this case, the rank of the piece increases by at most one. 
    See Figure~\ref{fig:BaseNode}(c). 
    We add two children of $t$, and assign the two pieces to those new nodes. Then we repeatedly subdivide the pieces further for the new nodes until all leaf noes have pieces containing exactly one edge of $H$. Then all pieces contains at most two vertices of $H$.
    
    Since $\wmax{H}=O(\log \ell)$ by Lemma~\ref{lem:sum-cost}, the new nodes constructed from a node $t$ of $T$ have weight at most $O(\log \ell)$ plus the weight of $t$, which is $O(\sqrt{\ell})$. This is because the closed polygonal curve $C$ constructed for non-unique maximal chains does not intersect any vertex of $H$ which were contained in the interior of $A_t$.
    In the case of a unique maximal chain, it intersects exactly one new vertex of $H$ which were contained in the interior of $A_t$.
    Since a single vertex has weight $O(\log \ell)$, this does not increase the weight of the surface decomposition asymptotically. 
    By the argument in this subsection, we can conclude that $(T, {\mathcal A})$ is a surface decomposition of $H$ of $O(\sqrt{\ell})$ weighted width.
	

	\begin{restatable}{lemma}{contracted}
	\label{lem:annuluswidth}
	$(T,  {\mathcal A})$ is a surface decomposition of $H$ and has weighted width $O(\sqrt{\ell})$.
	\end{restatable}

    \subsection{Step 3: sc-Decomposition of \texorpdfstring{$G$}{Lg} from a Decomposition of \texorpdfstring{$H$}{Lg}}\label{sec:step-3}

    Then we compute an sc-decomposition $(T, \bar {\mathcal A})$ of $G$ from the surface decomposition $(T,  {\mathcal A})$ of $H$.
    The pieces of ${\mathcal A}$ are constructed with respect to the vertices of $H$. 
    We perturb the boundaries of the pieces of ${\mathcal A}$ slightly so that no piece
    contains a vertex of $G$ on its boundary.     

    \subsubsection{Constructing an sc-Decomposition of \texorpdfstring{$G$}{Lg} Using Perturbation}
    For perturbation, observe that the pieces of all leaf nodes of $T$ is a partition of $\mathbb{R}^2$ by \textbf{(A2)}. 
    Thus if we perturb such pieces only, then the pieces of all other nodes of $T$ can be updated accordingly. 
    Let $\mathcal A_L$ be the set of the pieces of the leaf nodes of $T$.
    During the perturbation, we do not move the intersection points between $\partial A_t$ and the boundaries of the cells of $\mathcal M$, and thus we can handle the cells of $\mathcal M$ one by one separately. 
    
    For a cell $M$ of $\mathcal M$, let $\mathcal S_M$ be the subdivision of $M$ with respect to the pieces of $\mathcal A_L$. 
    Without changing its combinatorial structure, we modify $\mathcal S_M$ such that no 
    vertex of $G$ lies on the boundary of the edges of the subdivision.
    Since each cell of $M$ has $O(1)$ edges, we can do this without increasing the complexity of the subdivision. 
    By construction, each region of the resulting set $\mathcal A_L$ is a piece, and the combinatorial structure of the subdivision of $\mathbb{R}^2$ induced by the pieces of $\mathcal A_L$
    remains the same. 
    Let $\bar{ \mathcal A}$ be the set of pieces we obtained from the perturbation. 
    \begin{lemma} \label{lem:tcproperty}
        $(T, \bar{\mathcal A})$ is an sc-decomposition of $G$.
    \end{lemma}
    \begin{proof}
        By the general position assumption, no vertex of $G$ lies on the boundary of a cell of $\mathcal M$. Therefore,
        by construction, no vertex of $G$ lies on the boundary of a piece of $\bar{\mathcal A}$, and thus $V(G)\cap \partial \bar A_t=\emptyset$ (Condition \textbf{(C1)}). 
        Also, $\bar A_{t'}, \bar A_{t''}$ are interior-disjoint, and $\bar A_{t'}\cup \bar A_{t''}=\bar A_t$ 
        for all node $t$ and its children $t', t''$ (Condition \textbf{(C2)}). This is 
        because the combinatorial structures of the pieces of ${\mathcal A}$ do not change during the perturbation.
        For a leaf node $t$ of $T$, $A_t$ contains at most two vertices of $H$ by \textbf{(A3)}.
        For an $O(1)$-neighboring cell $M$ of the cell containing
        a vertex of $G$ of degree at least three, its boundary edges are contained in the drawing of $H$. Thus $M$ is intersected by $\bar A_t$ only when $A_t$ contains a vertex of $H$ lying on the boundary of $M$.
        Thus $A_t$ intersects at most two such cells.
        For the other cells, they contain at most two vertices of $G$.
        Thus $\bar A_t\cap V(G)$ are contained in at most two cells of $\mathcal M$, and thus the condition \textbf{(C3)} also holds.
        For the condition \textbf{(C4)},
        observe that a cell of $\mathcal M$ containing exactly 
        one point of $V(G)$ satisfies \textbf{(C4)} immediately.
        On the other hand, for a cell $M$ containing more than one point of $V(G)$,
        the boundary of $M$ appears on the drawing of $H$, and thus a piece containing a vertex of $V(G)\cap M$
        contains a vertex of $H$ on its boundary. Since the leaf pieces are interior-disjoint,
        there are $O(1)$ leaf nodes containing vertices of $V(G)\cap M$ in their pieces.
        Therefore, \textbf{(C4)} is also satisfied. 
    \end{proof}

    \subsubsection{Analysis of the Clique-Weighted Width} \label{sec:sc-width}
    We show that the clique-weighted width of $(T, \bar{\mathcal A})$ is $O(\sqrt \ell)$ using the fact that $(T,\mathcal A)$ has weighted width $O(\sqrt \ell)$.
    In particular, we show that for each node $t$ of $T$, the clique-weight of the 
    cells of $\mathcal M$ containing the endpoints of the edges of
    $\cut t$ is $O(\sqrt \ell)$.
    Let $A$ be a piece of a node of $T$, and let $\bar A$ be the piece obtained from $A$ by the perturbation.

    Let $t$ be a node of $T$. 
    Let $\bdr t$ denote the set of vertices of $V(H)\cap \partial A_t$.
    The weight of $t$ in the surface decomposition of $H$ is defined as the total weight of $\bdr t$.
    Also, the weight $c_H(v)$ of $v$ in $H$ is defined as the total clique-weight of the cells of $\mathcal M_\textsf{nb}(v)$, where $\mathcal M_\textsf{nb}(v)$ denotes the set of all $\alpha$-neighboring cells of $M_v$. 
    Therefore, it suffices to show that the endpoints of $\cut t$ are contained in the cells of 
    the union of $\mathcal M_\textsf{nb}(v)$ for all vertices $v\in \bdr t$. 
    


 Let $e=ab$ be an edge of $\cut t$, and let $M$ be a cell containing a point of $e\cap \partial \bar A_t$.
    Consider the case that either $M_a$ or $M_b$ contains a vertex of degree at least three. Here, $M_a$ and $M_b$ are the cells of $\mathcal M$ contain $a$ and $b$, respectively.
    Notice that $M$ is an $\alpha$-neighboring cell of a vertex of $G$ of degree at least three, and thus $\partial M$ appears on the drawing of $H$ by the construction of $H$.
    Therefore,
    $\partial A_t$ intersects $\partial M$ only at vertices of $H$, say $v$. 
    Note that $v\in \bd(t)$, and 
    $a$ and $b$ are contained in cells of $\mathcal M_\textsf{nb}(v)$.

    Now consider the case that $M_a$ and $M_b$ contain vertices
    of degree two only. 
    Thus $a$ and $b$ are degree-2 vertices of $G$.
    Consider a maximal chain $\pi$ of degree-2 vertices of $G$ containing $ab$.
    It is a part of the drawing of $H$, and thus $\partial A_t$ intersects $\pi$ 
    only at vertices of $\pi$.
    Therefore, $ab$ is intersected by $\partial A_t$ only when $a$ or $b$ lies on $\partial A_t$.
    Without loss of generality, assume that $a$ lies on $\partial A_t$.
    This means that $a\in \bdr t$, and both $a$ and $b$ are contained in cells of
    $\mathcal M_\textsf{nb}(a)$ in this case.

    \annuluscarving*

    \carvingudg*
    \subparagraph{Remark.}
    The \textbf{icf}-property is crucial to get the bound on the width of the surface decomposition.
    In particular, for a graph $G$ which does not admit this property, we cannot bound the number of vertices of $H_3$.
    Nevertheless, in this case, we can compute an sc-decomposition of weighted width $O(\sqrt{\ell+\ell'})$ once we have a map of $G$, where $\ell$ and $\ell'$ denotes the number of vertices of degree at least three in $G$ and $H$, respectively.

    \section{Properties of Vertex-Disjoint Cycles} \label{sec:property}
    For a graph $G$ drawn in the plane with a map that satisfies the \textbf{icf}-property, 
    we define three properties of a set of $k$ vertex-disjoint cycles in $G$: quasi-planar property, bounded packedness property, and sparse property. 
    We will see in Section~\ref{sec:dp}, if $G$ has $k$ vertex-disjoint cycles satisfying these three properties, we can compute a set of $k$ vertex-disjoint cycles in $G$ in $2^{O(\sqrt{k})}n^{O(1)}$ time. 
    Let $\mathcal M$ be a map, and $H$ be a map sparsifier of $G$ with respect to $\mathcal M$. 
    Throughout this section, $\alpha$ 
    is a fixed constant which is the maximum number of cells of $\mathcal M$ intersected by one edge of $G$.

	\subparagraph{Bounded packedness property.}
	A set $\Gamma$ of vertex-disjoint cycles of $G$ is \emph{$c$-packed} if 
	at most $c$ vertices contained in a cell of $\mathcal M$ lie on the cycles of $\Gamma'$, 
	and every cycle of $\Gamma\setminus \Gamma'$ is a triangle, 
 	where  $\Gamma'$ is the set of cycles of $\Gamma$ visiting at least two vertices from different cells 
        of $\mathcal M$. 

\begin{figure}
		\centering
		\includegraphics[width=0.5\textwidth]{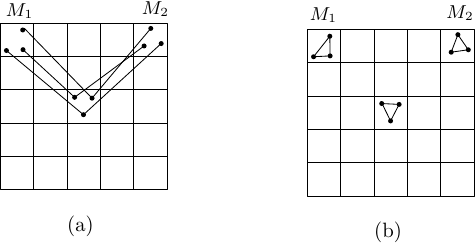}
		\caption{\small The $\alpha$-neighboring cells around $M$. 
		If there are three vertices $v_1,v_2$ and $v_3$ in $M$ such that the neighboring vertices of them on $\Gamma'$ are contained in $M_1$ and $M_2$, 
		the three cycles containing $v_1,v_2$ and $v_3$, respectively, can be replaced with the three triangles. 
		}
		\label{fig:packedness}
	\end{figure}

 
	\begin{restatable}{lemma}{packed}
	    \label{lem:packed-general}
         Let $G$ be a graph drawn in the plane 
         having a map $\mathcal M$.
         If $G$ has a set of $k$ vertex-disjoint cycles, then 
         it has a set of $k$ vertex-disjoint cycles which is $O(1)$-packed. 
	\end{restatable}
 \begin{proof}
		Among all sets of $k$ vertex-disjoint cycles of $G$, we choose the one $\Gamma$ that minimizes 
		the number of cycles of $\Gamma$ visiting at least two vertices from different cells. 
            Let $\Gamma'$ be the set of cycles of $\Gamma$  visiting at least vertices from different cells. 
		Let $M$ be a cell of $\mathcal M$, and 
		$Q$ be the set of vertices of $V\cap M$ lying on the cycles of $\Gamma'$. 
		In the following, we show that $|Q| \leq \beta^2 \times 3$, where $\beta$ is the maximum number of $\alpha$-neighboring cells of a cell of $\mathcal M$.
    		
		For this purpose, assume $|Q| > \beta^2 \times 3$. 
		For each vertex $v$ in $Q$, let $x(v)$ and $y(v)$ be the neighbors of $v$ along the cycle of 
		$\Gamma'$ containing $v$. 
		Consider the (at most $\beta$) cells of $\mathcal M$ 
        which are $\alpha$-neighboring cells of $M_v$.
		All neighbors of the vertices of $Q$ are contained in the union of these cells.  
		Among all pairs of such cells, there is a pair $(M_1, M_2)$ such that 
		the number of vertices $v$ of $Q$ with $x(v)\in M_1$ and $y(v)\in M_2$ is at least $|Q|/\beta^2\geq 3$. 
		See Figure~\ref{fig:packedness}(a--b).
		Let $v_1, v_2$ and $v_3$ be such vertices. Then we can replace the cycles containing at least one of $v_1, v_2$ and $v_3$
		with three triangles: one consisting of $x(v_1),x(v_2)$ and $x(v_3)$, one consisting of $v_1,v_2$ and $v_3$, and one consisting of $y(v_1),y(v_2)$ and $y(v_3)$. This is because the vertices of $G$ contained in a single cell form a clique, and the cycles of $\Gamma$ are pairwise vertex-disjoint.
		This contradicts the choice of $\Gamma$. 
	\end{proof}

   \subparagraph{Quasi-planar property.} 
A set $\Gamma$ of vertex-disjoint cycles of $G$ 
     is \emph{quasi-planar} if each cycle of $\Gamma$ is not self-crossing, and
     for any set $\Pi$ of subpaths of the non-triangle cycles of $\Gamma$, 
     the intersection graph of $\Pi$ 
     is $K_{z,z}$-free for a constant $z$ 
     depending only on the maximum number of paths of $\Pi$ containing a common edge. 
     Here, the intersection graph is defined as the graph where a vertex corresponds to a non-triangle cycle of $\Gamma$, and two vertices are connected by an edge if and only if their corresponding cycles cross in their drawing. 
     For two paths of $\Pi$ having a common edge $e$, their intersection at a point in $e$ is not counted as crossing. 
     In this case, note that two paths of $\Pi$ come from the same cycle of $\Gamma$. 

         \begin{figure}
		\centering
		\includegraphics[width=0.6\textwidth]{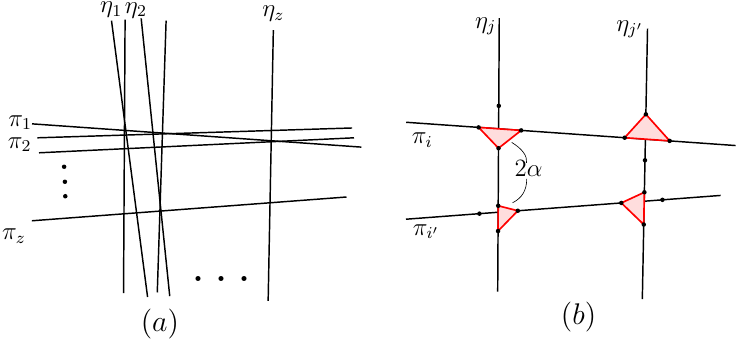}
		\caption{\small (a) The intersection graph of $\eta_1,\ldots, \eta_{z},\pi_1,\ldots, \pi_{z}$ contains $K_{z,z}$ as a subgraph, and none of two paths share a common edge.
        (b) We can pick four indices such that the the distance between two cells containing two crossing points is at least $2\alpha$. We can replace (at most) four cycles containing $\pi_i, \pi_{i'},\eta_j$ and $\eta_{j'}$ with four triangles by the \textbf{icf}-property.
		}
		\label{fig:quasi-planar}
	\end{figure}
 
\begin{lemma}\label{lem:pseudo-planar}
        Let $G$ be a graph drawn in the plane 
         having a map $\mathcal M$ and admitting the \textbf{icf}-property.
         If $G$ has a set of $k$ vertex-disjoint cycles, then 
         it has a set of $k$ quasi-planar and $O(1)$-packed vertex-disjoint cycles. 
	\end{lemma}
        \begin{proof}
        Let $\Gamma$ be a set of $k$ 
        vertex-disjoint cycles which has the minimum number of edges in total among all $O(1)$-packed vertex-disjoint cycles.        
        First, due to the \textbf{icf}-property, no cycle of $\Gamma$ has a self-crossing.
        Let $\Pi$ be the set of subpaths of non-triangle cycles of $\Gamma$ 
         such that two subpaths of $\Pi$ may share some common edges, and
        let $\mathcal I$ be the intersection graph of the drawing of $\Pi$. 
         Also, let $s$ be the the maximum number of paths of $\Pi$ containing a common edge. 

    If $\mathcal I$ is not $K_{z,z}$-free, there are 	
    $z/s$ paths $\pi_1,\ldots \pi_{z}$ and another $z/s$ paths $\eta_1,\ldots, \eta_{z}$ of 
    $\Pi$ such that 
    $\pi_i$ crosses $\eta_j$ in their drawings 
    for every index pair $(i,j)$ with $1\leq i,j\leq z$. 
    Let $a(i,j)$ be a crossing point between $\pi_i$ and $\eta_j$. 
    We denote the cell of $\mathcal M$ containing $a(i,j)$ by $M_{i,j}$.
    Let $\beta$, $\kappa$ be the maximum numbers of $\alpha$-neighboring cells and $2\alpha$-neighboring cells of a cell of $\mathcal M$, respectively. Then $\beta, \kappa=O(1)$.
    Since each edge of $G$ intersects at most $\alpha$ cells, 
    at most $b:=c^2\beta^2$ crossing points are contained in one cell $M$ of $\mathcal M$. Here, $c$ is a constant specified in Lemma~\ref{lem:packed-general} such that $G$ admits a $c$-packed solution.  
    We claim that if $z/s>2b^2\kappa^2$, there are four indices $i,j,i',j'<z$ such that 
    $d(M_{i,j},M_{i',j})$, $d(M_{i,j'},M_{i',j'})$, $d(M_{i,j},M_{i,j'})$, and $d(M_{i',j},M_{i',j'})$ are greater than $2\alpha$, 
    where $d(M,M')$ is the length of the shortest path between $M$ and $M'$ in the dual graph of $\mathcal M$.
    If the claim holds, we can replace the cycles containing
    $\pi_i,\pi_{i'},\eta_j$ and $\eta_{j'}$ into four vertex-disjoint triangles by applying \textbf{icf}-property on the crossing points $a(i,j), a(i',j), a(i,j')$ and $a(i',j')$.
    The four triangles are vertex-disjoint because the distance between any two cells containing two crossing points are greater than $2\alpha$, while each edge intersects at most $\alpha$ cells.
    Moreover, this replacement does not violate the $c$-packed property, but it violates the choice of $\Gamma$.
    Thus the lemma holds. 
    
    Then we show that we can always find four such indices. 
    Since $z/s>2b^2\kappa^2$, we can pick 
    $2b\kappa$ indices $j_1,\ldots j_{2b\kappa}$ such that 
    the distance between any two of $M_{1,j_1},\ldots M_{1,j_{2b\kappa}}$ is greater than $2\alpha$.
    Then there are at most $2b^2\kappa^2$ indices $i$ such that $d(M_{1,j_m}, M_{i, j_m})$ is at most $2\alpha$ for some $m\leq 2b\kappa$.
    In other words, we can pick an index $i$ such that $d(M_{1,j_m},M_{i,j_m})$ is greater than $2\alpha$ for all $m\leq 2b\kappa$.
    Among $2b\kappa$ intersection points $a(i,j_m)$'s, we can pick two points $a(i,j)$ and $a(i',j')$ such that $d(M_{i,j}, M_{i',j'})>2\alpha$.
    See also Figure~\ref{fig:quasi-planar}.     This completes the proof. 
    \end{proof}

    \subparagraph{Sparse property.} 
    A graph $G$ with the \textbf{icf}-property that has a map becomes sparse after applying the cleaning step that
    removes all vertices not contained in any cycle of $G$. The cleaning step works by recursively removing a vertex of degree at most one from $G$. Note that the resulting graph is also a graph with the \textbf{icf}-property. 
    From now on, we assume that every vertex of $G$ has degree at least two.
    
    \begin{restatable}{lemma}{numdegree-apd}
	    \label{lem:num-degree-apd}
     Let $G$ be a graph  with the \textbf{icf}-property.
    For a constant $c$, assume that $G$ has more than $ck$ vertices of degree at least three 
     after the cleaning step. Then $(G,k)$ is a \textbf{yes}-instance, and 
     moreover, we can find a set of $k$ vertex-disjoint cycles in polynomial time.
    \end{restatable}
    \begin{proof}   
    We first remove $O(k)$ vertices to make the graph planar. 
    If two edges $xy$ and $zw$ of $G$ cross, 
    three of four vertices, say $x,y$ and $z$, form a triangle by the \textbf{icf}-property.   
    Then we remove $x,y,z$ and all edges incident to $x,y,z$ from $G$. 
    If we can do this $k$ times, $G$ has $k$ vertex-disjoint cycles. 
    Thus we assume that we can apply the removal less than $k$ times. In this way, we remove $O(k)$ vertices of $G$, and then we have a planar induced subgraph $G_1$ of $G$. Note that $G_1$ itself also admits the \textbf{icf}-property. 
    We first show that the number of vertices of $G_1$ adjacent to the removed vertices is $O(k)$. 
    Observe that no five vertices of $G_1$ are contained in the same cell of $\mathcal M$. Otherwise, $G_1$ contains $K_5$ as a subgraph, and thus it is not planar. 
    For a removed vertex $v$, its neighbors in $G$ are contained in $\beta=O(1)$ cells of $\mathcal M$.
    Therefore, there are $O(1)$ vertices of $G_1$ adjacent to $v$ in $G$, and thus 
    the number of vertices of $G_1$ adjacent to removed vertices is $O(k)$. 

    Now it is sufficient to show that the number of vertices of $G_1$ of degree at least three is $O(k)$. 
    For the purpose of analysis, we iteratively remove degree-1 vertices from $G_1$. Let $G_2$ be the resulting graph. In this way, a vertex of degree at least three in $G_1$ can be removed, and the degree of a vertex can decrease due to the removed vertices.
    We claim that the number of such vertices is $O(k)$ in total. 
    To see this, consider the subgraph of $G_1$ induced by the vertices removed during the construction of $G_2$. They form a rooted forest 
    such that the root of each tree is adjacent to a vertex of $G_2$, and no leaf node is adjacent to a vertex of $G_2$. 
    The number of vertices of the forest of degree at least three is linear in the number of leaf nodes of the forest. 
    Since $G$ has no vertex of degree one due to the cleaning step,
    all leaf nodes of the forest are adjacent to $V(G)\setminus V(G_1)$.
    Therefore, only $O(k)$ vertices of degree at least three in $G$ are removed during the construction of $G_2$. 
    Also, the degree of a vertex decreases only when it is adjacent in $G$ to a root node of the forest.
    Since the number of root nodes of the forest is $O(k)$, and a root node of the forest is adjacent to only one vertex of $G_2$, there are $O(k)$ vertices of $G_2$ whose degrees decrease during the construction of $G_2$.

    Therefore, it is sufficient to show that vertices of $G_2$ of degree at least three is at most $10k$. 
    Let $H$ be the dual graph of $G_2$. Since $H$ is also a planar graph, it is 5-colorable.
    Then there is a set of  $|F_2|/5$ vertices with the same color, and it is an independent set of $H$,
    where $F_2$ denotes the number of faces of $G_2$. 
    Note that a vertex of $H$ corresponds to a face of $G_2$.
    Two vertices of $H$ are connected by an edge if and only if their corresponding faces share an edge. Therefore, the faces in $H$ corresponding to the vertices of $I$ form $|I|$ vertex-disjoint cycles in $G_2$. Therefore, if $|F_2|$ exceeds $5k$, $(G,k)$ is \textbf{yes}-instance. 
    In this case,
    since we can find a 5-coloring of a planar graph in polynomial time, we can compute $k$ vertex-disjoint cycles in polynomial time.
    Now consider the case that $(G,k)$ is \textbf{no}-instance. 
    Let $V_2, E_2$ and $F_2$ be the vertex set, edge set, and face set of $G_2$, respectively.
    Let $\ell$ be the number of vertices of $G_2$ of degree at least three. We have three formulas:
    \begin{align*}
        &2(|V_2|-\ell)+3\ell \leq \sum_{v\in V_2} \textsf{deg}(v) = 2|E_2|\\
        &|V_2|-|E_2|+|F_2|=2 \\
        &|F_2|\leq 5k
    \end{align*}
    This implies that $|V_2| \leq 10k$, and thus the lemma holds. 
\end{proof}

\section{Generalization of Catalan Bounds to Crossing Circular Arcs}\label{sec:catalan}
    It is well-known that, for a fixed set $P$ of $n$ points on a circle, the number of different sets of pairwise non-crossing circular arcs having their endpoints on $P$ is the $n$-th Catalan number, which is $2^{O(n)}$. 
    Here, two circular arcs are \emph{non-crossing} if 
    they are disjoint or one circular arc contains the other circular arc.
    This fact is one of main tools used in the ETH-tight algorithm for the planar cycle packing problem: 
    Given a noose $\gamma$ of a planar graph $G$ visiting $m$ vertices of $G$, the parts of the cycles of $\Gamma$ contained in the interior of $\gamma$ corresponds to 
    a set of pairwise non-crossing circular arcs having their endpoints on fixed $m$ points, 
    where $\Gamma$ denotes a set of vertex-disjoint cycles of $G$ crossing $\gamma$. 
    This holds since no two vertex-disjoint cycles cross in planar graphs. 
    Although two vertex-disjoint cycles in a graph $G$ admitting the \textbf{icf}-property can cross in their drawing, there exists a set of $k$ vertex-disjoint cycles of $G$ with the quasi-planar property due to Lemma~\ref{lem:pseudo-planar}.
    To make use of this property, we generalize the Catalan bound on non-crossing circular arcs to circular arcs which can cross.
    
    The \emph{circular arc crossing graph}, CAC graph in short, is a variation of the circular arc graph. 
    For a fixed set $P$ of points on a unit circle, consider a set $\mathcal C$ of circular arcs on the unit circle connecting two points of $P$ such that no arcs share their endpoints. 
    The CAC graph of $\mathcal C$ is defined as  the graph whose vertex corresponds to a circular arc of $\mathcal C$, and 
    two vertices are connected by an edge if and only if their corresponding circular arcs are crossing.
    Here, we say two arcs on the unit circle are \emph{crossing} if they are intersect and none of them contains the other arc. 
    Note that the CAC graph $G_\textsf{cac}$ of $\mathcal C$ depends only on the pairing $\mathcal P$ of the endpoints of the arcs of $\mathcal C$. 
    If we do not want to specify circular arcs, we simply say $G_\textsf{cac}$ is the CAC graph of $\mathcal P$. 
    With a slight abuse of term, we say a set $\mathcal C$ of circular arcs (or its pairing) is $K_{z,z}$-free if
    its CAC graph is $K_{z,z}$-free. Note that a set of pairwise non-crossing circular arcs is $K_{1,1}$-free since its CAC graph is edgeless.
    
    \begin{restatable}{lemma}{circulararc}
    \label{lem:circular-arc}
        The number of $K_{z,z}$-free sets of circular arcs over a fixed set $P$ of points is $2^{O_z(|P|)}$.
    \end{restatable}
    \begin{proof}
    For the convenience, we let $P=\{1,2,\ldots, 2m\}$ be the cyclic sequence of the elements of $\mathbb Z_{2m}$. 
    For a $K_{z,z}$-free set $\mathcal C$ of circular arcs, an arc of $\mathcal C$ with counterclockwise endpoint $a$ and clockwise endpoint $b$ for $\mathbb Z_{2m}$ is denoted by $[a,b]$.  
    We decompose the arcs of $\mathcal C$ into several \emph{layers} as follows. 
    In each iteration, we choose all arcs of $\mathcal C$ not contained in any other arcs of $\mathcal C$, and remove them from $\mathcal C$. 
    The \emph{level} of $\gamma$ is defined as the index of the iteration when $\gamma$ is removed. 

    \begin{figure}
        \centering
        \includegraphics[width=0.4\textwidth]{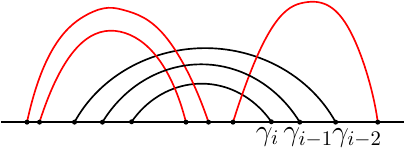}
        \caption{\small All arcs of level less than $i-z$ with endpoints in $\gamma_i$ (colored red) intersect $\gamma_i,\gamma_{i-1},\ldots,\gamma_{i-z+1}$.}
        \label{fig:CAC}
    \end{figure}
    \begin{claim}\label{claim:level}
        An arc of level $i$ contains at most $z$ endpoints of the arcs of level less than $i-z$.
    \end{claim}
    \begin{proof}
    Assume to the contrary that an level-$i$ arc $\gamma_i$ of $\mathcal C$ 
    contains more than $z$ endpoints of arcs of $\mathcal C$ of level less than $i-z$.     
    For each index $k$ with $i-z< k\leq i$, there is an level-$k$ arc $\gamma_k$ containing $\gamma_i$.
    This arc $\gamma_k$ forbids $\gamma_i$ from being assigned level $k$. 
    But $\gamma_k$ does not contain an arc $\gamma$ of level less than $i-z$ as $\gamma_k$ forbids $\gamma$ from being assigned level lower than $k$. 
    That is, $\gamma$ crosses $\gamma_k$ if $\gamma_i$ contains an endpoint of $\gamma$. See Figure~\ref{fig:CAC}. 
    Therefore, $\{\gamma_{i},\gamma_{i-1},\ldots \gamma_{i-z+1}\}$ and the $z$ arcs of $\mathcal C$ of level less than $i-z$ crossing $\gamma_i$ induces $K_{z,z}$ subgraph of the CAC graph of $\mathcal C$. This contradicts that the CAC graph of $\mathcal C$ is $K_{z,z}$-free.
    \end{proof} 
    
    Note that the maximum level among all arcs of $\mathcal C$ is at most $m$ since the endpoints of the arcs of $\mathcal C$ are distinct. 
    For each integer $i\leq m$, we denote the set of endpoints of all circular arcs of level at most $i$ by $P_i$. 
    Then we have $P_1\subset P_2\subset\ldots\subset P_m=P$.
    We let $P_{i}=\emptyset$ for $i\leq 0$ for convenience.
    For a subset $P'$ of $P$, we define a \emph{partial order} $\chi_{P'}:P'\rightarrow [1,|P'|]$ such that
    $\chi_{P'}(a)=b$ if $a$ is the $b$-th endpoint of $P'$ (starting from 1).
    We simply write $\chi_{i,j}=\chi_{P_j\setminus P_{i-1}}$
    for every $i\leq j$.
    Given a partial order $\chi_{i,j}$, consider the mapping that maps each circular arc $[a,b]$ into $[\chi_{i,j}(a), \chi_{i,j}(b)]$, where $a,b\in P_j\setminus P_{i-1}$. 
    We define the \emph{quotient} $Q_{i,j}$ by the set of circular arcs $[\chi_{i,j}(a), \chi_{i,j}(b)]$ on the ground set $[1,2,\ldots |P_j\setminus P_{i-1}|]$. 
    The CAC graph of $Q_{i,j}$ is isomorphic to the subgraph of the CAC graph of $\mathcal C$, and therefore it is $K_{z,z}$-free.

    \medskip
    Then we analyze the number of $K_{z,z}$-free sets of circular arcs over $|P|$. First, 
    the number of different tuples $(|P_1|,|P_2|,\ldots |P_m|)$ is $2^{O(m)}$ by simple combinatorics. Hence, we assume that the tuple $(|P_1|,|P_2|,\ldots |P_m|)$ is fixed. 
    We claim that the number of different quotients $Q_{1,i}$ under a fixed $Q_{1,i-1}$ is single exponential in $|P_i\setminus P_{i-z-1}|$ for every index $i$.
    First we show that the number of different $Q_{i,i}$ is $2^{O_z(|P_i\setminus P_{i-1}|)}$. 
    Observe that no arc of $Q_{i,i}$ contains the other arc of $Q_{i,i}$ by construction. For a circular arc $[a,b]$ in $Q_{i,i}$, at least $\frac{b-a}{2}$ circular arcs of $Q_{i,i}$ contain $a$, and these arcs induce a clique of size $\frac{b-a}{2}$ in the CAC graph of $Q_{i,j}$. Since this graph is $K_{2z}$-free,
    $b-a\leq 4z$. Consequently,  for each $a$, there are $4z$ different candidates for the other endpoint $b$ of $[a,b]$. 
    Thus, the number of different $Q_{i,i}$ is $2^{O_z(|P_i\setminus P_{i-1}|)}$.
   
    Next we bound the number of different $Q_{i-z, i}$.
    Note that the partial order $\chi_{i-z, i-1}$ is fixed since $Q_{1,i-1}$ is fixed.
    Hence, the number of different $\chi_{i-z, i}$ is equal to the number of ways to determine the position of 
    $P_i\setminus P_{i-1}$ on $\chi_{i-z, i}$ times the number of different quotient $Q_{i,i}$.
    The former one equals to the number of different subsets of size $|P_i\setminus P_{i-1}|$ from $\{1,2,\ldots |P_i\setminus P_{i-z}|\}$, which is $2^{O(|P_i\setminus P_{i-z}|)}$. Thus, the number of different $Q_{i-z, i}$ is $2^{O_z(|P_i\setminus P_{i-z}|)}$. 
    Finally we compute the number of different $Q_{1,i}$.
    We compute the number of different range $\chi_{1,i}(P_{i-z})$ using Claim~\ref{claim:level}.
    This is equal to the number of solutions of the following equation:
    \[n(1)+n(2)+\ldots+n(|P_i\setminus P_{i-z}|)=|P_{i-z}|, \text{ } \forall j, n(j)\in \mathbb Z_{\geq 0}.\]
    Here, $n(j)$ represents the number of endpoints of $P_{i-z}$ located between
    the $(j-1)$-th and $j$-th endpoints of $P_i\setminus P_{i-z}$. 
    We use the notation $a_j$ to denote the $j$-th endpoint of $P_i\setminus P_{i-z}$.
    Since $Q_{1,i-1}$ is fixed, $n(j)$ is uniquely determined if both $a_j$ and $a_{j-1}$ are contained in $P_{i-1}$.
    Therefore, we enough to compute $n(j)$ for all $j$ with $a_j\notin P_{i-1}$ or $a_{j-1}\notin P_{i-1}$.
    Suppose $a_j\notin P_{i-1}$, and consider the arc of level $i$ whose one endpoint is $a_j$. 
    Then either this arc contains $[a_{j-1}, a_j]$, or there is an arc of level $i-1$ contains $[a_{j-1}, a_j]$.
    For both cases, at most $O(1)$ points of $P_{i-z-1}$ are contained in $[a_{j-1}, a_j]$ due to Claim~\ref{claim:level}. In addition, there are at most $2|P_i\setminus P_{i-1}|$ $n(j)$'s which are not uniquely determined. 
    Overall, the number of different $\chi_{1,i}(P_{i-z})$ is equal to the number of solutions of the following equation:
    \[n'(1)+n'(2)+\ldots+n'(2|P_i\setminus P_{i-1}|)=|P_i\setminus P_{i-z-1}|+O(1)\cdot |P_i\setminus P_{i-1}|, \text{ } \forall j, n'(j)\in \mathbb Z_{\geq 0}.\]
    This is
    $
    2^{O(|P_i\setminus P_{i-1}|)+O(|P_i\setminus P_{i-z-1}|)}=2^{O(|P_i\setminus P_{i-z-1}|)}.
    $
    Overall, the number of different $Q_{1,i}$ is at most the number of different $\chi_{1,i}(P_{i-z})$ times the number of different $Q_{i-z, i}$, which is $2^{O_z(|P_i\setminus P_{i-z-1}|)}$. This confirms the claim.
    Finally, the number of different $K_{z,z}$-free sets of circular arcs is 
    \[\exp(\sum_i O_z(|P_i\setminus P_{i-z-1}|))=\exp(O_z(|P_m|))=\exp(O_z(|P|)).\]
    This completes the proof.
\end{proof}

    Moreover, we can compute all 
    $K_{z,z}$-free sets of circular arcs over a fixed set $P$ in $2^{O_{z}(|P|)}$ time.
    \begin{restatable}{lemma}{circulararc-algo}
    \label{lem:circular-arc-algo}
    Given a set $P$ of points on a unit circle, we can compute all $K_{z,z}$-free sets of circular arcs over $P$ in $2^{O_z(|P|)}$ time.
    \end{restatable}
    \begin{proof}
    Let $|P|=2m$.
    Since the number of different tuples $(|P_1|,|P_2|,\ldots,|P_m|)$ is $2^{O(m)}$, 
    it is sufficient to compute all $K_{z,z}$-free sets of circular arcs under a fixed tuple $(|P_1|,|P_2|,\ldots,|P_m|)$.
    Lemma~\ref{lem:circular-arc} implicitly says that the number of different quotients $Q_{i,j}$ is $\exp(\sum_{k\in [i,j]}|P_k|)$.
    We show that given all different quotients $Q_{1,i-1}$, we can compute all different quotients $Q_{1,i}$ in $\exp(O(\sum_{j\leq i}|P_j|)\cdot |P|^2$ time.
    First, we enumerate all different quotients $Q_{i,i}$ in $2^{O(|P_i|)}$ time by exhaustively considering all cases that the range of $Q_{i,i}$ consists of the circular arcs of length at most $4z$.
    The number of different pairs $(Q_{1,i-1}, Q_{i,i})$ at this moment is $\exp(\sum_{j\leq i}|P_j|)$.
    For each different pair, we consider $\exp(\sum_{j\leq i}|P_j|)$ possible sets of $\chi_{1,i}(P_i)$.
    Each triple $(Q_{1,i-1}, Q_{i,i}, \chi_{1,i}(P_i))$ induces a unique quotient $Q_{1,i}$. Then we check whether the CAC graph w.r.t. the resulting quotient $Q_{1,i}$ is $K_{z,z}$-free or not, in $O(|P|^2)$ time for each triple.
    In summary, we can compute all different quotients $Q_{1,i}$ in $\exp(O(\sum_{j\leq i}|P_j|))|P|^2$ time.
    Then we can compute all different quotients $Q_{1,m}$ in $\exp(O(\sum_{j\leq m}|P_j|))|P|^3=2^{O(|P|)}$ time.
    \end{proof}

    \section{Algorithm for \cpp}\label{sec:dp}
    In this section, given a graph $G=(V,E)$ drawn in the plane 
    along with a map $\mathcal M$ and an sc-decomposition $(T,\mathcal A)$ of clique-weighted width $\omega$, 
    we present a $2^{O(\omega)}n^{O(1)}$-time dynamic programming algorithm that computes a maximum
    number of vertex-disjoint cycles of $G$.

    Throughout this section,
    let $\Gamma$ be a maximum-sized set of vertex-disjoint cycles which is quasi-planar and $O(1)$-packed. 
    Let $\finerm$ be the finer subdivision of $\mathcal M$ formed by subdividing each cell of $\mathcal M$ into \emph{finer cells} along $\partial A$ for all pieces $A$ corresponding to the leaf nodes of $T$. 
    We call a cycle of $\Gamma$ an \emph{intra-cell} cycle if its vertices are contained in a single finer cell of $\finerm$,
    and an \emph{inter-cell} cycle, otherwise.
    Notice that an intra-cell cycle is a triangle since $\Gamma$ is $O(1)$-packed. 
    To make the description easier, 
    we sometimes consider a set of vertex-disjoint cycles and paths of $G$ 
    as a \emph{graph} consisting of disjoint paths (path components) and cycles (cycle components) if it is clear from the context. 
    For a set $\Gamma'$ of paths and cycles of $G$, we let $\ev {\Gamma'}$ be the set of end vertices of the \emph{paths} of $\Gamma'$,
    and $V(\Gamma')$ be the set of all vertices of \emph{paths and cycles} of $\Gamma'$.

    \subsection{Standard DP Algorithm for Cycle Packing} \label{sec:standard-dp}
    In this subsection, we present a $2^{O(\omega \log \omega)}$-time algorithm for the cycle packing problem using a dynamic programming on $(T,\mathcal A)$ in a standard manner. 
    For a node $t$ of $T$, recall that $\cut t$ is the set of edges of $E$ with at least one endpoints in $A_t$ that intersect the boundary of $A_t$. 
    Also, recall that $G_t$ is the subgraph of $G$ induced by $V\cap A_t$.
    Let $\mathcal M_t$ be the set of finer cells of $\finerm$ having vertices of $V(\cut t)\cap A_t$. 
    The clique-weight of the finer cells in $\mathcal M_t$ is $O(\omega)$ as 
    for each cell $M$ of $\mathcal M$, $V(G)\cap M$ is decomposed into $O(1)$ subsets $V(G)\cap M_\textsf{f}$
    for finer cells $M_\textsf{f}$ of $\finerm$ 
     by \textbf{(C4)}. 
    
    

\begin{figure}
	\centering
	\includegraphics[width=0.45\textwidth]{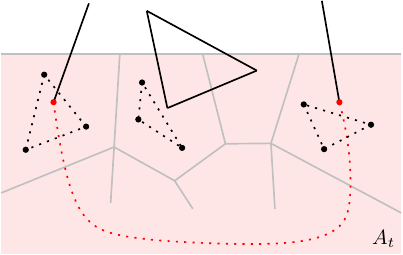}
	\caption{\small 
	   The $t$-signature $(\Lambda, V(\Delta), \mathcal P)$ consists of the vertex-disjoint cycles and paths induced by $\cut t$ (black lines), the vertices contained in $\mathcal M_t$ forming intra-cell cycles of $\Delta$ (black vertices), and the connectivity information of the endpoints of the paths of $\Lambda$ (dotted red line). 
	}\label{fig:dp-standard1}
\end{figure}

    \subparagraph{Subproblems.}
    To encode the interaction between $\Gamma$ and $\cut t$, we focus on the parts of $\Gamma$ consisting of edges of $\cut t$.
    Let $\Lambda$ be the subgraph of $\Gamma$ induced by $\cut t$, and $\Delta$ be the set of intra-cell cycles of $\Gamma$ contained in a cell of $\mathcal M_t$. See Figure~\ref{fig:dp-standard1}. 
    Then $V(\Lambda)\cap V(\Delta)=\emptyset$, since
     an edge of $\Lambda$ is a part of an inter-cell cycle of $\Gamma$.
    
    The interaction between $\Gamma$ and $\cut t$ can be characterized as
	$(\Lambda, V(\Delta), \mathcal P)$ where 
    $\Lambda$ and $V(\Delta)$ are defined earlier, 
    and $\mathcal P$ is the pairing of the vertices of $\ev {\Lambda}$ contained in $A_t$ such that the 
    vertices in the same pair belong to the same component of the subgraph of $\Gamma$ induced by $V\cap A_t$.
    We call $(\Lambda, V(\Delta), \mathcal P)$ the $t$-\emph{signature} of $\Gamma$. 
    Consider two sets $\Gamma$ and $\Gamma'$ of vertex-disjoint cycles of $G$ having the same signature. 
    Imagine that we replace the parts of the cycles of $\Gamma$  contained in $G_t$ with the
    parts of the cycles of $\Gamma'$ contained in $G_t$. 
    The resulting set forms vertex-disjoint cycles of $G$. 
    Therefore, it suffices to find a maximum number of vertex-disjoint cycles and paths contained in $G_t$ that matches the given signature. A formal proof will be given in Lemma~\ref{lem:dp-correct}. 

    Since we are not given $\Gamma$ in advance, we try all possible tuples which can be the signatures of $\Gamma$, and find an optimal solution for each of such signatures. 
    We define the subproblem for each \emph{valid} tuple $(\widetilde{\Lambda}, 
    \widetilde{V}, \widetilde{\mathcal P})$. 
    We say a tuple $(\widetilde{\Lambda}, 
    \widetilde{V}, \widetilde{\mathcal P})$ is \emph{valid} if for every finer cell $M$ of $\mathcal M_t$
    
    \begin{itemize}\setlength\itemsep{-0.1em}
        \item $\widetilde{\Lambda}$ 
        induces vertex-disjoint paths and cycles consisting of edges of $\cut t$,
        \item $\widetilde{V}$ is 
        a subset of $\{v\in M'  \mid M'\in \mathcal M_t\}$ with  $\widetilde V\cap V(\widetilde \Lambda
        )=\emptyset$ and 
        $|M \cap (V\setminus \widetilde V)|=O(1)$, and 
	    \item $\widetilde{\mathcal P}$ is a pairing of all vertices of $\ev{\widetilde{\Lambda}
     }$ contained in $A_t$.
    \end{itemize}

    Notice that the $t$-signature of $\Gamma$ is also a valid tuple. 
    For each valid tuple $Q$ for a node $t$, 
    the value $S[Q]$ is defined as 
    the maximum number of vertex-disjoint cycles of $\Gamma'$ contained in $G_t$ over all sets $\Gamma'$ of vertex-disjoint cycles and paths of $G[E(G_t)\cup \cut t]$ whose $t$-signature is $Q$.\footnote{The signature is defined for a set of disjoint cycles of $G$, but it can be extended to a set of disjoint paths and cycles of $G[E(G_t)\cup \cut t]$ in a straightforward manner.}
    If such a set does not exist, we let $S[Q]=-\infty$.

	\begin{restatable}{lemma}{numtuple}\label{lem:num-tuples}
		The number of valid tuples is $2^{O(\omega\log \omega)}$. 
	\end{restatable}
 \begin{proof}
        First, we analyze the number of choices of $\widetilde\Lambda$. 
        Observe that every finer cell of $\mathcal M_t$ contains $O(1)$ vertices of $\widetilde\Lambda$ 
        due the second validity condition.         
        Since the clique-weight of $V(\cut t)$ is $O(\omega)$ by the definition of the clique-weight of an sc-decomposition, 
        the number of different sets of $V(\widetilde \Lambda
        )$ satisfying the validity conditions is 
        \[\prod_{M\in\mathcal M_t} |M\cap V(\cut t)|^{O(1)} = \exp\biggl( O \biggl( \sum_{M\in\mathcal M_t} \log (|M\cap V(\cut t)|+1)\biggr)\biggr) = 2^{O(\omega)}.\]
        Since each vertex of $\widetilde \Lambda
        $ has degree at most two in $\widetilde \Lambda
        $, the number of different edge sets of $\widetilde \Lambda
        $ is $2^{O(\omega)}$.
        Second, the number of different $\widetilde V$ is $2^{O(\omega)}$ because all but $O(1)$ vertices in $M$ are contained in $\widetilde V$ for each finer cell $M\in\mathcal M_t$, and the clique-weight of the finer cells of $\mathcal M_t$ is $O(\omega)$.
        Thus similarly to the first case, we can show that the number of different choices of $\w V$ is  $2^{O(\omega)}$. 
        Finally, observe that the number of vertices of $\ev{\widetilde \Lambda}
        $ contained in $A_t$ is $O(\omega)$ since 
        the number of cells of $\mathcal M_t$ containing vertices of $\widetilde \Lambda
        $ is 
        $O(\omega)$.
        Then the number of different pairings $\mathcal P$ is $2^{O(\omega \log \omega)}$. 
	\end{proof}



    \begin{figure}
        \centering
        \includegraphics[width=0.7\textwidth]{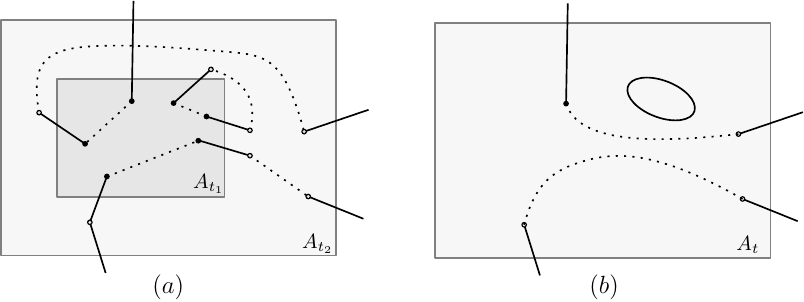}
        \caption{\small (a) For the two tuples $Q_1, Q_2$ harmonic with $Q$, the edge set of $\w \Lambda_u$ consists of the black lines of $\w \Lambda_1\cup \w \Lambda_2$ and dotted curves of $\w{\mathcal P}_1\cup \w{\mathcal P}_2$.
        (b) $\w \Lambda_u$ consists of one cycle and two paths, which are stored as the information of $c(Q_1,Q_2)$ and $\w{\mathcal P}$, respectively.}
        \label{fig:dp-relation}
    \end{figure}
    
    \subparagraph{Relation of the subproblems.}	
    For a valid tuple $Q=(\widetilde {\Lambda}, \widetilde{V}, \widetilde{\mathcal P})$ for a node $t$, 
    we say two valid tuples $Q_1= (\widetilde \Lambda_1, \widetilde V_1, \widetilde{\mathcal P}_1)$ and $Q_2= (\widetilde \Lambda_2, \widetilde V_2, \widetilde{\mathcal P}_2)$ for the two children $t_1$ and $t_2$ of $t$, respectively, 
    are \emph{harmonic} with $Q$ 
    if the following four conditions hold. 
    Let $\widetilde \Lambda_\textsf{u}$ be the graph obtained by taking  
    the union of $\w{\Lambda}_1$ and $\w{\Lambda}_2$ and then 
    by adding the edges $uv$ for all subsets $\{u,v\}\in (\widetilde{\mathcal P}_1\cup \widetilde{\mathcal P}_2)$.
    See Figure~\ref{fig:dp-relation}. 
    Then the following are the four conditions for tuples $(Q_1,Q_2)$ harmonic with $Q$.
    \begin{itemize}\setlength\itemsep{-0.1em}
        \item $M\cap \widetilde V=M\cap \widetilde V_i$ for every finer cell $M$ of $\mathcal M_{t_i}\cap \mathcal M_t$ for $i=1,2$, 
        \item $E(\w{\Lambda}_1)\cap \cut {t_2} = E(\w{\Lambda}_2)\cap \cut {t_1}$  and $E(\w{\Lambda}_i)\cap \cut t = E(\w{\Lambda})\cap \cut {t_i}$ for $i=1,2$,
        \item $\widetilde \Lambda_\textsf{u}$ consists of vertex-disjoint paths and cycles, and 
        \item $\widetilde {\mathcal P}$ is the pairing consisting of the endpoint pairs of paths of $\widetilde \Lambda_\textsf{u}$.
    \end{itemize}
    
    Notice that for fixed two tuples $Q_1$ and $Q_2$ harmonic with $Q$, the graph $\widetilde\Lambda_\textsf{u}$ is fixed. 
    Let $c(Q_1,Q_2)$ be the number of cycles in $\widetilde\Lambda_\textsf{u}$. 

    \begin{lemma}\label{lem:dp-correct}
	For a valid tuple $Q=(\widetilde {\Lambda}, \widetilde{V}, \widetilde{\mathcal P})$, 
	we have 
	\[
	S[Q] =  \max\{S[Q_1]+S[Q_2]
         + c(Q_1,Q_2) \}
	\]
	where in the maximum we consider harmonic tuples $(Q_1,Q_2)$ with $Q$. 
	\end{lemma}
	\begin{proof} 
    The value $S[Q]$ is defined as the maximum number of vertex-disjoint cycles of $\Gamma'$ contained in $G_t$ among all sets $\Gamma'$ of vertex-disjoint cycles and paths in $G[E(G_t)\cup \cut t]$ whose $t$-signature is $Q$. 
    We first show that $S[Q]\leq \max\{S[Q_1]+S[Q_2] + c(Q_1,Q_2)\}$. 
    Among all sets of vertex-disjoint paths and cycles we can attain for $S[Q]$, we choose the one $\Gamma'$ which is $O(1)$-packed and quasi-planar.\footnote{The argument of Section~\ref{sec:property} implies that if there is a set of vertex-disjoint paths and cycles whose $t$-signature is $Q$, then there is a set of vertex-disjoint paths and cycles whose $t$-signature is $Q$ and which is $O(1)$-packed and quasi-planar.}
    Let $t_1$ and $t_2$ be the two children of $t$. 
    Let $Q_i=(\Lambda_i, V(\Delta_i), \mathcal P_i)$ be the $t_i$-signature of $\Gamma'$ for $i=1,2$.
    Since the signatures of $O(1)$-packed and quasi-planar sets of vertex-disjoint cycles/paths are all valid, 
     $Q_1$ and $Q_2$ are also valid. Also, it is not difficult to see that $Q_1$ and $Q_2$ are harmonic with $Q$.

    We analyze $S[Q]$ with respect to $S[Q_1]$, $S[Q_2]$, and $c(Q_1,Q_2)$. 
    Recall that the number of cycles of $\Gamma'$ contained in $G_t$ is $S[Q]$. 
    The cycles of $\Gamma'$ contained in $G_t$ can be partitioned into three cases: it is contained in $G_{{t}_1}$, or contained in $G_{{t}_2}$, or intersects both. 
    The number of cycles of the first case is at most $S[Q_1]$, and the number of cycles of the second case is at most $S[Q_2]$. 
    A cycle $\Gamma'$ of the third case is an inter-cell cycle. Imagine that we remove $\cut {t_1}\cup \cut {t_2}$ from $\Gamma'$.
    Then we have a number of paths connecting two vertices of $\ev{\cut {t_1}\cup \cut {t_2}}$ each of which is contained in $G_i$ for $i=1,2$.
    Therefore, each path corresponds to a pair of $\mathcal P_i$, and thus a cycle of the third case corresponds to a cycle of $\Lambda_\textsf{u}$.
    The number of cycles of the third case is at most $c(Q_1,Q_2)$, and thus the inequality holds.

    Now consider the other direction: $S[Q]\geq \max\{S[Q_1]+S[Q_2] + c(Q_1,Q_2)\}$.
    For this, let $(Q_1, Q_2)$ be a pair of tuples harmonic with $Q$. Let $\Gamma'_1$ and $\Gamma'_2$ be the sets of paths and cycles we attain for $S[Q_1]$ and $S[Q_2]$, respectively, which are quasi-planar and $O(1)$-packed. Also, let $Q_i=(\Lambda_i, V_i, \mathcal P_i)$.
    Let $\Gamma'$ be the union of $\Gamma_1'$ and $\Gamma_2'$. Since $(Q_1,Q_2)$ is a tuple harmonic with $Q$, 
    the subgraph of the union of $\Gamma'$ induced by $E[G_{t_i}]\cup \cut {t_i}$ is exactly $\Gamma_i'$. 
    Also, it is a set of disjoint paths and cycles whose signature is $Q$. Therefore,  $\Gamma'$ is considered
    in the computation of $S[Q]$, and thus the number of cycles of  $\Gamma'$ contained in $G_t$ is at most $S[Q]$. 

    Thus it is sufficient to analyze the number of cycles of $\Gamma'$ with respect to $S[Q_1]$, $S[Q_2]$, and $c(Q_1,Q_2)$.   
    Let $\Lambda_\textsf{u}$ of the union of $\Lambda_1, \Lambda_2$ and the edges $uv$ with $\{u,v\}\in\mathcal P_1\cup\mathcal P_2$. 
    We replace each edge $uv$ of $\Lambda_\textsf{u}$ with $\{u,v\}\in\mathcal P_i$ with the path in $\Gamma_i'$ between $u$ and $v$.
    It is not difficult to see that the resulting set $\Lambda_\textsf{u}$ is the union of $\Gamma'_1$ and $\Gamma'_2$ excluding the cycles  contained in $G_{t_i}$ for $i=1,2$. 
    The number of cycles  contained in $G_{t_i}$ is $S[Q_i]$ for $i=1,2$.
    A cycle intersecting both $G_{t_1}$ and $G_{t_2}$ contains a cut edge of $\cut {t_1}\cap \cut {t_2}$, and it is 
    a cycle of  $\Lambda_\textsf{u}$, and thus the number of such cycles is $c(Q_1,Q_2)$.
    Therefore, the inequality holds.    
	\end{proof}

	Then we can compute $S[Q]$ for all valid tuples $Q$ in $2^{O(\omega\log \omega)} n^{O(1)}$ time in total,
	and thus we can solve \cp in  $2^{O(\omega\log \omega)} n^{O(1)}$ time. Here, the bound on $2^{O(\omega\log \omega)}$ comes from
 the number of different pairings $\mathcal P$, and thus in the following section, we focus on decreasing the number of different pairings $\mathcal P$ using the $O(1)$-packed property and quasi-planar property of $\Gamma$. 


    \subsection{Analysis of Crossing Patterns}\label{sec:handling-crossing-edges-apd}
    In this subsection, we introduce a key idea of our improvement on the standard DP algorithm. To do this, we analyze the crossing pattern of inter-cell cycles of $\Gamma$ more carefully. 
    The subgraph of $\Gamma$ induced by $V\cap A_t$ consists of paths and cycles, and each path has both endpoints in $V(\cut t)$. 
    Let $\Pi$ be the set of path components of the subgraph of $\Gamma$ induced by $V\cap A_t$. 
    Notice that a path of $\Pi$ is part of an inter-cell cycle of $\Gamma$ since intra-cell cycles contain no edge of  $V(\cut t)$.  
    For a path $\pi$ of $\Pi$, 
    the path obtained from $\pi$ by adding the two cut edges incident to $\pi$ at the end of $\pi$ is denoted by $\dvec{\pi}$. Notice that it is possible that $\dvec{\pi}$ is a cycle.

    In the following, we define the ordering of paths $\pi$ of $\Pi$ with respect to a crossing point between $\dvec{\pi}$ and $\partial A_t$. 
    For this purpose, let $\bar \pi$ be the path (polygonal curve) obtained from $\dvec{\pi}$ by removing parts of its end edges maximally 
    such that the endpoints of $\bar \pi$ lie on $\partial A_t$. 
    We call $\bar \pi$ the \emph{anchored path} of $\pi$. 
    Let $\bar \Pi$ be the set of the anchored paths of the paths of $\Pi$. 
    A path of $\bar \Pi$ is fully contained in $A_t$ by the definition of $\cut t$.
    
   
    \subparagraph{Simple case: $\partial A_t$ is a single curve.}
     Given a fixed set $P$ of $O(\omega)$ vertices on $\partial A_t$ (which is indeed $\ev{\bar \Pi}$), our goal is to compute a small number of pairings of $P$ 
    one of which is a correct pairing of $\bar \Pi$. 
    We first focus on the simple case that
    all points of $\ev{\bar \Pi}$ lie on a single boundary curve $C$ of $A_t$. 
    We can compute
    $2^{O(\omega)}$ pairings one of which is a correct pairing as follows. 
    For a path $\bar \pi$ of $\bar \Pi$, we represent it as the pair of its end vertices on $C$. 
    Then the circular arc crossing graph of all pairs of $\bar\Pi$ is isomorphic to a subgraph of the intersection graph of $\Pi$, and thus it is $K_{z,z}$-free by the quasi-planar property. 
    That is, for any two paths $\bar\pi$ and $\bar\pi'$ of $\bar\Pi$, one ending at $a$ and $b$, and one ending at $a'$ and $b'$, if $a,a',b$ and $b'$ appear along $C$ in this order, the two paths must cross as they cannot cross $\partial A_t$. 
    Using this, we simply enumerate all pairings of $P$ whose corresponding circular arc graphs are $K_{z,z}$-free by Lemma~\ref{lem:circular-arc-algo}. The number of such pairings is $2^{O(|P|)}=2^{O(\omega)}$, and we can compute them in $2^{O(\omega)}$ time. 
    \subparagraph{General case: $\partial A_t$ has more than one curves.}
    Now we consider the general case where two endpoints of $\bar \pi$ are contained in different boundary curves of $\partial A_t$.
    In particular, let $(C,C')$ be a pair of
     boundary curves of $A_t$. 
    To handle the paths $\bar\pi_{x,y}$ of $\bar\Pi$ with $x\in C$ and $y\in C'$, we connect $C$ and $C'$ with an arbitrary curve $\lambda=\lambda(C,C')$. This curve intersects $C$ and $C'$ only at its endpoints and does not contain any endpoints of the paths of $\bar\Pi$. 
    Subsequently, we obtain
    the closed curve consisting of $C, C'$ and $\lambda$ such that $\lambda$ appears in the closed curve twice.
    Since none of the endpoints of a path in $\bar\Pi$ lie on $\lambda$, we can orderly arrange the endpoints of paths in $\bar\Pi$ such that each path has one endpoint on $C$ and the other on $C'$. This arrangement follows along the described closed curve, which consists of $\lambda$, $C$, and $C'$.
    
\begin{figure}
		\centering
		\includegraphics[width=0.8\textwidth]{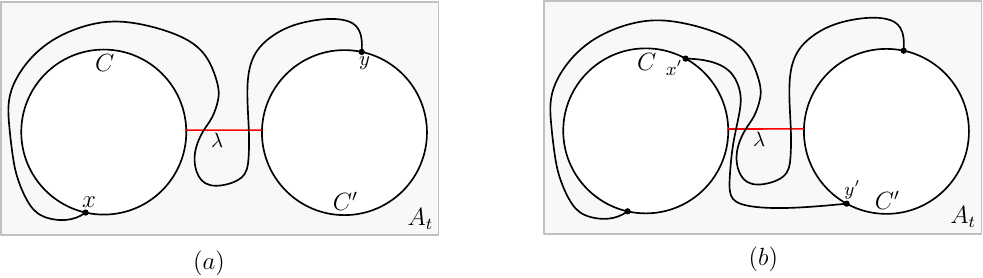}
		\caption{\small (a) The union of $C,C'$ and $\lambda$ (red curve) forms a closed curve. The path $\bar \pi_{x,y}$ having crossing number 2.
		(b) Illustrates two paths whose four endpoints are cross-ordered. Their crossing numbers differ in modulo 2.
        }
		\label{fig:crossing-number}
    \end{figure}

    We will show that for two paths $\bar\pi_{x,y}$ and $\bar\pi_{x',y'}$ with $x,x'\in C$ and $y,y'\in C'$, 
    if they are cross-ordered, and their \emph{crossing numbers} are the same in modulo 2, then  $\bar \pi_{x,y}$ and $\bar \pi_{x',y'}$ cross. Here, 
    the crossing number of a path $\bar\pi_{x,y}$ with $x\in C$ and $y\in C'$ is defined as the number of times that $\bar \pi_{x,y}$ crosses $\lambda(C,C')$.
    Note that $\bar \pi_{x,y}, \bar \pi_{x',y'}$ never cross either $C$ or $C'$.
    A single edge of $\bar\pi_{x,y}$ and $\bar \pi_{x',y'}$ can cross $\lambda(C,C')$ multiple times, and each crossing point is counted as one. See also figure~\ref{fig:crossing-number}.
    Then we have the following lemma. Its proof is deferred to Section~\ref{sec:proof-crossed-order}.

    \begin{lemma}\label{lem:crossed-order}
        For any two paths $\bar\pi_{x,y}$ and $\bar\pi_{x',y'}$ with $x,x'\in C$ and $y,y'\in C'$ for two boundary curves $C$ and $C'$ of $A_t$, 
        if they are cross-ordered, and their crossing numbers are the same in modulo 2, then $\bar \pi_{x,y}$ and $\bar \pi_{x',y'}$ cross. 
    \end{lemma}

    Therefore, the circular arc crossing graph of the pairing of the paths $\bar\pi_{x,y}$ of $\bar\Pi$ with $x\in C$ and $y\in C'$ having the same crossing numbers 
    is $K_{z,z}$-free for a constant $z$.

    \subsection{Improved Algorithm by Considering Crossing Patterns} 
 \label{sec:improvedboundcp}
  In this subsection, we encode the interaction between $\Gamma$ and $\cut t$ along with the crossing patterns.
  This encoding is characterized by
	$(\Lambda,V(\Delta), \mathcal P, h)$ where  
    $\Lambda, V(\Delta)$ and $\mathcal P$ are defined earlier, 
    and
    $h$ is the mapping that maps each path of $\Pi$ to the parity of its crossing number. 
    We call $(\Lambda,V(\Delta),\mathcal P, h)$ the \emph{refined $t$-signature} of $\Gamma$. 
    Since we are not given $\Gamma$ in advance, we try all possible tuples which can be the signatures of $\Gamma$, and find an optimal solution for each of such signatures. That is, 
    we define the subproblem for each \emph{valid} tuple $Q=(\w \Lambda, \w V,\w {\mathcal P}, \w h)$ as follows.
    For each cell $M$ of $\mathcal M_t$, 
    
    \begin{itemize}\setlength\itemsep{-0.1em}
        \item $\widetilde{\Lambda}$ induces vertex-disjoint paths and cycles consisting of edges of $\cut t$,
        \item $\widetilde{V}$ is 
        a subset of $\{v\in M' \mid M'\in \mathcal M_t\}$ with  $\widetilde V\cap V(\widetilde \Lambda)=\emptyset$ and $|(M\cap (V\setminus  \widetilde V)|=O(1)$, 
	\item $\widetilde{\mathcal P}$ is a pairing of all vertices of $\ev{\widetilde{\Lambda}}$ contained in $A_t$ such that its 
circular arc crossing graph with respect to $\w h$ is $K_{z,z}$-free, and
    \item $\w h$ is a mapping that maps each pair of $\w {\mathcal P}$ to $\{\textsf{even},\textsf{odd}\}$.
    \end{itemize}

    The circular arc crossing graph of $\widetilde P$ with respect to $\w h$ is defined as follows.  First, we choose the first crossing point $x(a)$ between $\partial A_t$ and $e(a)$ starting from $a$ for each vertex $a\in \ev{\w \Lambda}\cap A_t$, where $e(a)$ denotes the edge of $\Lambda$ having endpoint $a$. 
    We consider a pair $(C, C')$ of boundary curves of $A_t$ as a unified closed curve $\delta(C, C')$, employing the same technique as previously by inserting a single curve $\lambda(C, C')$ twice.
    In this way, we obtain $O(1)$ closed curves $\delta(C,C')$.
    Here, $C$ and $C'$ can be the same. 
    For a boundary curve $C$ of $A_t$, we construct the circular arc crossing graph of all pairs $(a,b)$ with $x(a)\in C$ and $x(b)\in C$. 
    For a pair $(C,C')$ of boundary curves of $A_t$, we construct two circular arc crossing graphs, one for the pairs $(a,b)$ of $\w P$
    with $x(a)\in C$, $x(b)\in C'$ and $\w h(a,b)=\textsf{even}$, and one for the pairs $(a,b)$ of $\w P$
    with $x(a)\in C$, $x(b)\in C'$ and $\w h(a,b)=\textsf{odd}$.
    Then the circular arc crossing graph of $\w P$ respect to $\w h$ is the union of all such circular arc crossing graphs.

    \medskip 
    For a valid tuple $Q=(\w \Lambda, \w V, \w {\mathcal P},\w h )$, we let $S[Q]$ be 
    the maximum number of vertex-disjoint cycles of $\Gamma'$ contained in $G_t$ over all sets $\Gamma'$ of vertex-disjoint cycles and paths of $G[E(G_t)\cup \cut t]$ whose signature is $Q$. 
    If such cycles and paths does not exists, we let $S[Q]=-\infty$.
    By Lemma~\ref{lem:crossed-order} together with our previous arguments,
    the refined $t$-signature of $\Gamma$ is also valid.
    Therefore, we can obtain an optimal solution once we have $S[Q]$ for all valid tuples $Q$.
    
    \begin{lemma}
        The number of valid tuples is $2^{O(\omega)}n^{O(1)}$. 
    \end{lemma}
    \begin{proof}
        The number of choices of $\w \Lambda$ and $\w V$ is $2^{O(\omega)}$ as the same as the standard DP.
        To analyze the number of choices of $\mathcal P$, note that $x(a)$ is fixed for all points $a\in \ev{\w \Lambda}$ once we have $\w \Lambda, \w V$. 
        For each point $x(a)$, we guess the boundary curve of $\Gamma$ where its counterpart, say $y(a)$, lies. The total number of such choices is $2^{O(\omega)}$.
        For each pair $(C,C')$ of boundary curves of $\Gamma$ and for each $p\in\{\textsf{even},\textsf{odd}\}$,
        let $P(C,C',p)$ be the set of points $x(a)$ with $a\in\ev{\w \Lambda}$ such that $x(a)\in C$, $y(a)\in C'$, and $\w h(a)=p$.
        Then the possible pairings $\mathcal P$ are the combinations of all pairings of $P(C,C',p)$ which are $K_{z,z}$-free for all choices $(C,C',p)$. 
        The number of such pairings of $P(C,C',p)$ is $2^{O(\omega)}$ by Lemma~\ref{lem:circular-arc}.
        Since the number of choices of $(C,C',p)$ is $O(1)$, the total number of combinations of $\mathcal P$ is $2^{O(\omega)}$. 
        Finally, since the number of nodes of $T$ is $n^{O(1)}$, the number of all valid tuples is $2^{O(\omega)}n^{O(1)}$. 
    \end{proof} 

    Since the $t$-refined signature of $\Gamma$ contains all information of the original $t$-signature of $\Gamma$,
    we can give a relation between subproblems similar to the standard DP algorithm of Section~\ref{sec:standard-dp}.
    For a valid tuple $Q=(\widetilde {\Lambda}, \widetilde{V}, \widetilde{\mathcal P},\cdot )$ and a node $t$, 
    we say two valid tuples $Q_1= (\widetilde \Lambda_1, \widetilde V_1, \widetilde{\mathcal P}_1,\cdot )$ and $Q_2= (\widetilde \Lambda_2, \widetilde V_2,  \widetilde{\mathcal P}_2, \cdot)$ for the two children $t_1$ and $t_2$ of $t$, respectively, 
    are \emph{harmonic} with $Q$ 
    if the following four conditions hold. 
    Let $\widetilde \Lambda_\textsf{u}$ be the graph obtained by taking  
    the union of $\w{\Lambda}_1$ and $\w{\Lambda}_2$ and then 
    by adding the edges $uv$ for all subsets $\{u,v\}\in (\widetilde{\mathcal P}_1\cup \widetilde{\mathcal P}_2)$.
    Then the following are the four conditions for tuples $(Q_1,Q_2)$ harmonic with $Q$. 
    
    \begin{itemize}\setlength\itemsep{-0.1em}
        \item $M\cap \widetilde V=M\cap \widetilde V_i$ for every cell $M$ of $\mathcal M_{t_i}\cap \mathcal M_t$ for $i=1,2$, 
        \item $E(\w{\Lambda}_1)\cap \cut{t_2} = E(\w{\Lambda}_2)\cap \cut{t_1}$  and $E(\w{\Lambda}_i)\cap \cut{t} = E(\w{\Lambda})\cap \cut{t_i}$ for $i=1,2$,
        \item $\widetilde \Lambda_\textsf{u}$ consists of vertex-disjoint paths and cycles, and 
        \item $\widetilde {\mathcal P}$ is the pairing consisting of the endpoint pairs of paths of $\widetilde \Lambda_\textsf{u}$.
    \end{itemize}

    Note that each tuple $Q$ contains additional information beyond $\w \Lambda, \w V$ and $\w{\mathcal P}$. 
    In fact, we encode this to enumerate $2^{O(\omega)}$ candidates for $\mathcal P$, and thus once $\mathcal P$ is fixed,
    we do not use $\w h$ in dynamic programming. 
    For fixed tuples $Q_1$ and $Q_2$ that are harmonic with $Q$, the graph $\widetilde\Lambda_\textsf{u}$ is uniquely determined. 
    We denote the number of cycles in $\widetilde\Lambda_\textsf{u}$ by $c(Q_1,Q_2)$.
    Then we can compute all $S[Q]$ in time polynomial in the number of valid tuples. 
    Since the following lemma can be proved in a similar manner to  Lemma~\ref{lem:dp-correct}, we omit its  proof.
    
    \begin{lemma}\label{lem:improve-dp-correct}
	For a valid tuple $Q=(\widetilde {\Lambda}, \widetilde{V},\widetilde{\mathcal P}, \w h)$, 
	we have 
	\[
	S[Q] =  \max\{S[Q_1]+S[Q_2]
         + c(Q_1,Q_2) \}
	\]
	where in the maximum we consider harmonic tuples $Q_1,Q_2$ over $Q$ with $Q_i= (\widetilde \Lambda_i, \widetilde V_i,\widetilde{\mathcal P}_i,\cdot)$. 
	\end{lemma}

    The following theorem summarizes this section. 
\begin{theorem}\label{thm:cp}
    For a graph with the \textbf{icf}-property given along with 
    an sc-decomposition of $G$ of clique-weighted width $\omega$ and a map of $G$ are given,
    \cp can be solved in $2^{O(\omega)}n^{O(1)}$ time.
\end{theorem}

\begin{theorem}
    Given a graph with the \textbf{icf}-property along with a map of $G$, 
    \cp can be solved in $2^{O(\sqrt k)}n^{O(1)}$ time.
\end{theorem}
\begin{proof}
    First, we compute a map sparsifier of $G$ with respect to the given map $\mathcal M$ in $n^{O(1)}$ time.
    As a preprocessing, we  recursively remove a vertex of degree at most one from $G$.     
    By Lemma~\ref{lem:num-degree-apd}, after this cleaning step, 
    if $G$ has more than $ck$ vertices of degree at least three for a constant $c$,
    $(G,k)$ is a \textbf{yes}-instance.
    Thus, if the number of vertices of degree at least three in $G$ is larger than $ck$,
    we apply Lemma~\ref{lem:num-degree-apd} to compute a set of $k$ vertex-disjoint cycles in polynomial time.
    Otherwise, we can compute an sc-decomposition of clique-weighted width $O(\sqrt k)$ in polynomial time by Theorem~\ref{thm:annuluscarving}.
    By Theorem~\ref{thm:cp}, we can solve the problem in  $2^{O(\sqrt k)}n^{O(1)}$ time.
\end{proof}

By Corollary~\ref{thm:carvingudg} and by the fact that a unit disk graph admits a map, the following holds. 
\begin{corollary}
    Given a unit disk graph $G$ with its geometric representation and an integer $k$,
    we can compute a set of $k$ vertex-disjoint cycles of $G$ in  $2^{O(\sqrt k)}n^{O(1)}$ time if it exists.
\end{corollary}

    \subsection{Proof of Lemma~\ref{lem:crossed-order}}\label{sec:proof-crossed-order}
\begin{figure}
        \centering
        \includegraphics[width=0.6\textwidth]{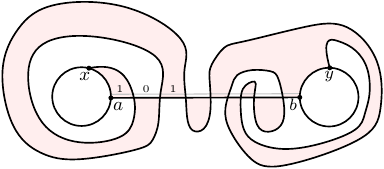}
        \caption{\small The gray curve $\lambda'$ lies close to the $a$-$b$ curve $\lambda$. Starting from 1, we assign the parity for the points of $\lambda$. The faces of $\mathcal F$ of parity one are colored red.}
        \label{fig:parity}
    \end{figure}

    Assume to the contrary that $\bar \pi_{x,y}$ and $\bar \pi_{x',y'}$ do not cross.
    In the following, we consider  $\bar \pi_{x,y}$ and $\bar \pi_{x',y'}$ are curves in the plane. 
    Let $a=\lambda\cap C$ and $b=\lambda\cap C'$ be two endpoints of $\lambda$. 
    Let $S$ be the closed curve defined by $C$, $C'$ and $\lambda$. 
    Moreover, let $U$ be the face of $\mathbb R^2\setminus S$ containing $A_t$.
    Then $\bar\pi_{x,y}$ divides $U$ into a set $\mathcal F$ of faces.
    Let $S_x$ (and $S_y$) be the $a\rightarrow x$ (and $b\rightarrow y$) subcurve of $C$ in the counterclockwise direction  (and the clockwise direction). 
    Let $F_x$ and $F_y$ be the faces of $\mathcal F$ whose boundaries contain $S_x$ and $S_y$, respectively.
    We choose two curves $\lambda'$ and $\lambda''$ connecting $C$ and $C'$ 
    such that they lie sufficiently close to $\lambda$ in the opposite directions: $\lambda'$ lies in $F_x$ and $\lambda''$ lies in $F_y$. 
    Let $a', b'$ be the endpoints of $\lambda'$ with $a'\in C$ and $b'\in C'$, and
    let $a'', b''$ be the endpoints of $\lambda''$ with $a''\in C$ and $b''\in C'$.
    We define the \emph{parity} of the points of $\lambda'\cup \lambda''$ which is either zero or one. 
    Imagine that we traverse $a'\rightarrow b'$ along $\lambda'$. 
    We assign parity zero to $a'$ and all points we encounter during the traversal until we encounter an intersection point of $\lambda'$ and $\bar\pi_{x,y}$.
    Then we assign parity one to the intersection point, and then we traverse further. Specifically, 
    we change the parity whenever the traversal encounters the intersection of $\lambda'$ and $\bar\pi_{x,y}$. 
    On the contrary, we assign parity one to $a''$, and then we change the parity whenever the traversal along $\lambda''$ encounters the intersection with $\bar\pi_{x,y}$. See Figure~\ref{fig:parity}. 

    We say the \emph{parity} of the face of $\mathcal F$ is one if there is a point $p$ of $\lambda' \cup \lambda''$ with parity one, and zero if there is a point of $\lambda' \cup \lambda''$ of parity zero.
    First we show that the parity of each face of $\mathcal F$ is well-defined.
    Let $p_1,p_2,\ldots p_r$ be the sequence of the intersection points between $\lambda$ and $\bar\pi_{x,y}$ along $\lambda$.
    We denote the counterpart of $p_i$ in $\lambda'$ (and $\lambda'')$ by $p'_i$ (and $p''_i$).
    The \emph{direction} of $p_i$ is \emph{down} if $p'_i$ appears immediately before $p''_i$ along the $x\rightarrow y$ traversal of $\lambda$. We say the direction of $p_i$ as \emph{up} otherwise.
    Let $p_i$ and $p_j$ be two consecutive intersection points between $\lambda$ and $\bar\pi_{x,y}$ along $\bar\pi_{x,y}$. Suppose $p_i$ appears immediately before $p_j$ along the $x\rightarrow y$ traversal on $\lambda$.
    We consider the region $D$ bounded by the union of the $p_i$-$p_j$ subcurve of $\bar\pi_{x,y}$ and the $p_i$-$p_j$ subcurve $\rho$ of $\lambda$.
    We claim that $i$ and $j$ are different in modulo 2 through the case analysis on $D$.
    \begin{claim}
        Two indices $i$ and $j$ are different in modulo 2.
    \end{claim}
    \begin{proof}
        If $D$ contains neither $a$ nor $b$ in its interior, $D\cap \bar\pi_{x,y}$ consists of disjoint curves whose endpoints are in $\rho$. Then the number of the intersection points in $\rho$ is even.
    Suppose $D$ contains not $y$ but $x$, and the direction of $p_i$ is up. Then the direction of $p_j$ is also up.
    In this case, either $p''_i\in D$ and $p'_j\notin D$, or $p''_i\notin D$ and $p'_j\in D$.
    Then the number of intersection points between $\rho$ and the $x$-$p''_i$ subcurve of $\bar\pi_{x,y}$ is even in the former case and odd in the latter case. Similarly, the number of intersection points between $\rho$ and the $p'_j$-$y$ subcurve of $\bar\pi_{x,y}$ is even in the former case and odd in the latter case. Then the number of intersection points in $\rho$ with $\bar\pi_{x,y}$ is even.
    Finally, if $D$ contains both $a$ and $b$, we may assume that $p_i$ goes up and $p_j$ goes down.
    In this case, either $p''_i, p''_j\in D$ or $p''_i,p''_j\notin D$.
    Similar to the previous case, we can show that $\rho$ contains a even number of intersection points. 
    \end{proof}
    
    Then we show that the parity of a face $F$ of $\mathcal F$ is well-defined. Since $\bar\pi_{x,y}$ is a simple curve and intersects $S$ only at their endpoints, each face $F$ is assigned to a parity.
    We claim that $F$ cannot contain both of a point of parity zero and a point of parity one. 
    Note that $F\cap \lambda'$ (and $F\cap \lambda''$) consists of $p'_i$-$p'_{i+1}$ (and $p''_j$-$p''_{j+1}$) subcurves of $\lambda'$ (and $\lambda''$) for indices $i$ (and $j$).
    Suppose $F$ contains the $p'_{i_1}$-$p'_{i_1+1}$ subcurve and the $p'_{i_2}$-$p'_{i_2+1}$ subcurve of $\lambda'$. 
    By the iterative use of the claim, $i_1$ and $i_2+1$ are different in modulo 2. Then $i_1$ and $i_2$ are same in modulo 2. 
    Similarly, if $F$ contains the $p''_{j_1}$-$p''_{j_1+1}$ subcurve and the $p''_{j_2}$-$p''_{j_2+1}$ subcurve of $\lambda''$, $j_1$ and $j_2$ are same in modulo 2.
    Next we consider the case that $F$ contains the $p'_i$-$p'_{i+1}$ subcurve of $\lambda$ and a $p''_j$-$p''_{j+1}$ subcurve of $\lambda''$.
    This happens only if $F$ contains the $p_a$-$p_b$ subcurve of $\bar\pi_{x,y}$ such that the directions of $p_a$ and $p_b$ are the same. Then either $F$ contains the $p'_a$-$p'_{a+1}$ subcurve and the $p''_b$-$p''_{b+1}$ subcurve, or the $p'_{a-1}$-$p'_a$ subcurve and the $p''_{b-1}$-$p''_b$ subcurve. In both cases, $a$ and $b$ are different in modulo 2 due to the claim. Since both ($i$, $a+1$) and ($j$, $b+1$) are pairs of intergers same in modulo 2, $i$ and $j$ are different in modulo 2. Then the parity of $F$ is well-defined.
    
    For the $x'\rightarrow y'$ traversal along $\bar\pi_{x',y'}$, the parity of a face we encounter changes whenever we cross $\lambda$ since $\bar\pi_{x,y}$ and $\bar\pi_{x',y'}$ do not cross.
    Since their crossing number is the same in modulo 2, 
    $x'$ is contained in $\partial F_x$ if and only if $y'$ is contained in $\partial F_y$. 
    This implies that the cyclic order of four points are either $\langle x',x,y,y'\rangle$ or $\langle x,x',y',y\rangle$. This contradicts that
    they are cross-ordered.

\bibliographystyle{plain}
\bibliography{paper}

\newpage
\appendix

\section{Applications of Surface Cut Decomposition}\label{sec:application}
In this section, we present algorithms for the non-parameterzied version of the \ocp, and the parameterzied versions of \dcp{} and \dbd{2} problems on unit disk graphs.
Let $\mathcal M$ be a map of $G$, which is a partition of the plane consisting of the interior-disjoint squares of diameter one.
Recall that
an sc-decomposition of $G$ is a pair $(T, \mathcal A)$ 
where $T$ is a rooted binary tree, and $\mathcal A$ is a mapping that maps a node $t$ of $T$ into a piece $A_t$ satisfying 
\textbf{(C1--C4)}.
For a node $t$ of $T$ and two children $t',t''$ and a cell $M$ of $\mathcal M$,

     \begin{itemize}\setlength\itemsep{-0.1em}
        \item \textbf{(C1)} $V(G)\cap \partial A_t=\emptyset$, 
	    \item \textbf{(C2)}  $A_{t'}, A_{t''}$ are interior-disjoint, $A_{t'}\cup A_{t''}=A_t$, 
        \item \textbf{(C3)} $V(G)\cap A_t$ is contained in the union of at most two cells of $\mathcal M$ for a leaf node $t$ of $T$, 
        \item \textbf{(C4)} 
    there are $O(1)$ leaf nodes of $T$ containing points of $M\cap V(G)$ in their pieces. 
    \end{itemize}

We restate the following statement from Section~\ref{sec:annuluscarving}.
\begin{restatable}{corollary}{appendixacdecomp} \label{lem:appendixacdecomp}
        Let $G=(V,E)$ be a unit disk graph and $\ell$ be the number of vertices of degree at least three in $G$.
        If the geometrc representation of $G$ is given, one can compute an sc-decomposition of clique-weighted width $\sqrt \ell$ in polynomial time. 
\end{restatable}

\subsection{Parameterized Algorithm of \dbd{2}}
We present a parameterized algorithm for the \dbd{2} problem on unit disk graphs running in time $2^{O(\sqrt k)}n^{O(1)}$. In this problem, we are given a unit disk graph $G$ and the solution size $k$ as a parameter,
then the objective is to find a subset $S$ of $V(G)$ of size at most $k$ such that 
the removal of $S$ make the remaining graph have a maximum degree exactly two.
We show the \emph{sparse property} for this problem: if $(G,k)$ is a \textbf{yes}-instance, then the number of vertices of $G$ of degree at least three is $O(k)$.
Let $S$ be a solution of size $k$ for the instance $(G,k)$.
Since $G-S$ has maximum degree two, the number of vertices of $G$ of degree at least three is at most the number of neighboring vertices of $S$ in $G-S$ plus $|S|$.
The neighbors of a vertex $v\in S$ in $G$ is contained in the union of $O(1)$ cells of $\mathcal M$. If the number of neighboring vertices of $v$ in $G-S$ is greater than $3\cdot O(1)$, there is a cell of $\mathcal M$ contains at least four vertices of $G-S$. Then the degree of each such vertex is at least three since they form a clique in $G-S$. Therefore, the number of neighboring vertices of $v$ in $G-S$ is $O(1)$. Now the number of vertices of degree at least three in $G$ is $k\cdot O(1)+k=O(k)$.
Due to the sparse property, we can compute an sc-decomposition of clique-weighted width $O(\sqrt k)$ in polynomial time. 

\subparagraph{Dynamic programming.} Then we apply the dynamic programming approach similar to the one outlined in Section~\ref{sec:dp}. Due to Observation~\ref{obs:udg-cross}, $G-S$ is a planar graph. Considering that $G-S$ consists of vertex disjoint paths and cycles, the \emph{quasi-planar} property holds for $G-S$: all cycles and paths are not self-crossing, and the intersection graph induced by these cycles and paths is $K_{z,z}$-free for $z=2$. 
Moreover, the proof of sparse property implicitly says that $G-S$ is $O(1)$-packed: each cell of $\mathcal M$ contains $O(1)$ vertices of $G-S$. 
As the running time of DP algorithm outlined in Section~\ref{sec:dp} is affected by both the bounded packedness property and the quasi-planar property, 
the resulting algorithm runs in time $2^{O(\sqrt k)}n^{O(1)}$.

\begin{theorem}
    Given an instance $(G,k)$ of \dbd{2} on unit disk graphs,
    we can can solve the problem in time  $2^{O(\sqrt k)}n^{O(1)}$.
\end{theorem}

\subsection{Parameterized Algorithm of \texorpdfstring{\dcp{}}{Lg}}
In this section, We present a parameterized algorithm for the \dcp{} problem on unit disk graphs running in time $2^{O_d(\sqrt k)}n^{O_d(1)}$. Given a unit disk graph $G$ and an integer $k$, the objective is to compute $k$ vertex disjoint $d$-cycles. 
We first apply the \emph{cleaning step}: for each vertex $v$ of $G$, we remove $v$ from $G$ if there is no $d$-cycle of $G$ containing $v$. 
We complete the cleaning step by searching all $O(n^d)$ candidate $d$-cycles of $G$. After that, we establish the \emph{sparse property} for this problem: if $(G,k)$ is a \textbf{no}-instance, then after the cleaning step, the number of vertices of $G$ of degree at least three is $O_d(k)$.
Let $S$ be the set containing the maximum number of vertex disjoint $d$-cycles of $G$, and 
$V(S)$ be the set of vertices contained in $S$. 
Each cell of $\mathcal M$ contains at most $d-1$ vertices of $G\setminus V(S)$ because $d$ vertices on the same cell forms a $d$-cycle. 
For each vertex $v$ of $G\setminus V(S)$, there is a $d$-cycle of $G$ containing $v$, but there is no such a cycle in $G\setminus V(S)$. Therefore, there is a vertex $u\in V(S)$ such that $v$ and $u$ are contained in a same $d$-cycle of $G$. Then the Euclidean distance $|uv|$ is $O_d(1)$.
Since $|V(S)|=dk$, the vertices of $G\setminus V(S)$ is contained in the union of $dk\cdot O_d(1)=O_d(k)$ cells of $\mathcal M$. 
Since each cell contains at most $d-1$ vertices of $G\setminus V(S)$, $|G\setminus V(S)|$ is $O_d(k)$. 
Thus,
the number of vertices of degree at least three in $G$ is at most $O_d(k)+dk=O_d(k)$.
Due to the sparse property, we can compute an sc-decomposition of clique-weighted width $O_d(\sqrt k)$ in polynomial time. 

\subparagraph{Dynamic programming.}
Then we apply the dynamic programming approach, similar to the one outlined in Section~\ref{sec:standard-dp}. For the sc-decomposition $(T,\mathcal A)$ of $G$ and a node $t$ of $T$, we define the subproblem for each valid tuple $(\widetilde{\Lambda}, 
    \widetilde{V}, \widetilde{\mathcal P}, \widetilde L)$.
    We say a tuple $(\widetilde{\Lambda}, 
    \widetilde{V}, \widetilde{\mathcal P}, \widetilde L)$ is \emph{valid} if for every cell $M$ of $\mathcal M_t$
    
    \begin{itemize}\setlength\itemsep{-0.1em}
        \item $\widetilde{\Lambda}$ 
        induces vertex-disjoint paths and cycles consisting of edges of $\cut t$,
        \item $\widetilde{V}$ is 
        a subset of $\{v\in M'  \mid M'\in \mathcal M_t\}$ with  $\widetilde V\cap V(\widetilde \Lambda
        )=\emptyset$ and 
        $|M \cap (V\setminus \widetilde V)|=O(1)$,  
	    \item $\widetilde{\mathcal P}$ is a pairing of all vertices of $\ev{\widetilde{\Lambda}
     }$ contained in $A_t$, and
     \item $\widetilde L$ is a function that maps the pairings of $\widetilde{\mathcal P}$ into $\{1,2,\ldots, d\}$.
    \end{itemize}
    
Intuitively, the length of the partial paths are encoded as $\widetilde L$. 
Similar to Section~\ref{sec:property}, we define the bounded-packedness property for $d$-cycles:

\begin{itemize}
    \item If $G$ admits $k$ vertex disjoint $d$-cycles, then $G$ has $k$ vertex disjoint $d$-cycles $\Delta$ such that each cell of $\mathcal M$ contains at most $O_d(1)$ vertices of $\Delta'$ where $\Delta'$ is the set of $d$-cycles of $\Delta$ visiting at least two vertices from different cells of $\mathcal M$.
\end{itemize}

The proof of this property follows the outline presented in the proof of Lemma~\ref{lem:packed-general}, and we omit the duplicated proof here. 
Then we show that the number of valid tuple is single exponential in $\sqrt k$.
 First, we analyze the number of choices of $\widetilde\Lambda$.
        Observe that every cell of $\mathcal M$ contains $O(1)$ vertices of $\widetilde\Lambda$ 
        due the $O(1)$-packedness.
        Since the clique-weight of $V(\cut t)$ is $O_d(\sqrt k)$ by the definition,
        the number of different sets of $V(\widetilde \Lambda
        )$ satisfying the validity conditions is 
        \[\prod_{M\in\mathcal M_t} |M\cap V(\cut t)|^{O(1)} = \exp\biggl( O \biggl( \sum_{M\in\mathcal M_t} \log (|M\cap V(\cut t)|+1)\biggr)\biggr) = 2^{O_d(\sqrt k)}.\]
        Since each vertex of $\widetilde \Lambda
        $ has degree at most two in $\widetilde \Lambda
        $, the number of different edge sets of $\widetilde \Lambda
        $ is $2^{O_d(\sqrt k)}$.
        Second, the number of different $\widetilde V$ is $2^{O_d(\sqrt k)}$ because all but $O(1)$ vertices in $M$ are contained in $\widetilde V$ for each cell $M\in\mathcal M_t$, and the clique-weight of the cells of $\mathcal M_t$ is $O_d(\sqrt k)$.
        Thus similarly to the first case, we can show that the number of different choices of $\w V$ is  $2^{O_d(\sqrt k)}$. 
        Third, for each pairing $(u,v)$ of $\w{\mathcal P}$, $|uv|$ is $O_d(1)$ as they contained in a same $d$-cycle.
        Due to the $O(1)$-packedness, the number of candidates $v$ to being $(u,v)\in \w{\mathcal P}$ is $O_d(1)$. 
        Then the number of different pairings $\mathcal P$ is $2^{O_d(\sqrt k)}$. 
        Finally, since the size of $\mathcal P$ is $O_d(\sqrt k)$, the number of different functions $\w L$ is $2^{O_d(\sqrt k)}$. 
        Overall, the number of subproblems is $2^{O_d(\sqrt k)}n^{O(1)}$. 

        The relation between subproblems almost follows the outline presented in Section~\ref{sec:standard-dp}. The only difference is that the harmonicity between two tuples now include a condition about the length of the combined cycles. This condition can be handled by the information stored in $\w L$. We skip the detailed analysis on the relation between subproblems.

\begin{theorem}
    Given an instance $(G,k)$ of \dcp{} on unit disk graphs,
    we can can solve the problem in time  $2^{O_d(\sqrt k)}n^{O_d(1)}$.
\end{theorem}



\subsection{Non-Parameterized Algorithm of \ocp{}}
In this section, We present an algorithm for the \ocp{} problem on unit disk graphs running in time  $2^{O(\sqrt n)}$. In this problem, the objective is to compute the set of maximum number of vertex-disjoint cycles of odd length. 
We compute an sc-decomposition of clique-weighted width $O(\sqrt n)$ by directly use Corollary~\ref{lem:appendixacdecomp}. 
Similar to Section~\ref{sec:property}, we define the bounded-packedness property for odd cycles: 

\begin{itemize}
    \item If $G$ admits $k$ vertex disjoint odd cycles, then $G$ has $k$ vertex disjoint odd cycles $\Delta$ such that each cell of $\mathcal M$ contains at most $O(1)$ vertices of $\Delta'$ where $\Delta'$ is the set of odd cycles of $\Delta$ visiting at least two vertices from different cells of $\mathcal M$.
\end{itemize}

The proof of this property follows the outline presented in the proof of Lemma~\ref{lem:packed-general}, and we omit the duplicated proof here. 

\subparagraph{Dynamic programming.} Then we apply the dynamic programming approach, similar to the one outlined in Section~\ref{sec:dp}. 
Let $\Delta$ be the set of vertex disjoint odd cycles.
Using the same approach with \dcp{} problem, we establish
\emph{quasi-planar} property for $\Delta$: all cycles are not self-crossing, and the intersection 
graph of the non-triangle cycles of $\Delta$ is $K_{z,z}$-free for a constant $z$.

A proof of the quasi-planar property follows the outline presented in the proof of Lemma~\ref{lem:pseudo-planar}. In particular, the proof of Lemma~\ref{lem:pseudo-planar} mainly shows that if the intersection graph has a subgraph isomorphic to $K_{z,z}$, then one can replace four vertex-disjoint odd cycles into four vertex-disjoint triangles. Since a triangle is an odd cycle of length 3, this implicitly shows that the intersection graph of the non-triangle cycles of $\Delta$ is $K_{z,z}$-free.
As the running time of DP algorithm outlined in Section~\ref{sec:dp} is affected by both the bounded packedness property and the quasi-planar property, applying the same approach gives a dynamic programming algorithm with the running time $2^{O(\sqrt n)}$.

\begin{theorem}
    Given a unit disk graph $G$,
    we can can solve \ocp{} on $G$ in time  $2^{O(\sqrt n)}$.
\end{theorem}

\end{document}